\documentclass[12pt]{article}
\usepackage{amssymb}
\makeatletter

\@addtoreset{equation}{section}
\def\section{\@startsection {section}{1}{\z@}{-2.5ex plus -1ex minus
 -.2ex}{1.3ex plus .2ex}{\large\bf}}
\def\subsection{\@startsection{subsection}{2}{\z@}{-2.25ex plus%
 -1ex minus -.2ex}{0.5ex plus .2ex}{\bf}}

\advance \voffset by -1.2in
\advance \hoffset by -0.6in
\def\Rea{\mbox{Re}}
\def\Ima{\mbox{Im}}

\def\cd{\!\cdot\!}

\def\Lor{L_3^\uparrow}
\def\LLor{\tilde L_3^\uparrow}
\def\LL{L}
\def\MM{M}
\def\ba{{\mbox{\boldmath $a$}}}
\def\bx{{\mbox{\boldmath $x$}}}
\def\by{{\mbox{\boldmath $y$}}}
\def\bj{{\mbox{\boldmath $j$}}}
\def\bk{{\mbox{\boldmath $k$}}}
\def\bp{{\mbox{\boldmath $p$}}}
\def\bP{{\mbox{\boldmath $P$}}}
\def\bq{{\mbox{\boldmath $q$}}}

\def\bP{{\mbox{\boldmath $P$}}}

\def\bob{{\mbox{\boldmath $b$}}}

\newcommand{\NN}{\mathbb{N}}
\newcommand{\ZZ}{\mathbb{Z}}

\newcommand{\RR}{\mathbb{R}}
\newcommand{\CC}{\mathbb{C}}
\newcommand{\tr}{{\rm tr}}
\def\bea{\begin{eqnarray}}
\def\eea{\end{eqnarray}}

\begin{document}
\parskip 6pt
\begin{flushright}
HWM-01-45\\
EMPG-02-07\\
ITFA-2002-12\\
hep-th/0205021
\end{flushright}

\begin{center}
\baselineskip 24 pt
{\Large \bf Quantum group symmetry and particle scattering  } \\
 {\Large \bf  in (2+1)-dimensional quantum gravity}

\baselineskip 18pt

\vspace{1cm}
{\large F.~A.~ Bais\footnote{\tt  bais@science.uva.nl},\,
       N.~M.~  Muller\footnote{\tt  nathalie.muller@worldtalk.cc} \\
Institute for Theoretical Physics, University of Amsterdam, \\
       Valckenierstraat 65, 1018 XE Amsterdam, The Netherlands\\
and \\
   B.~J.~Schroers\footnote{\tt bernd@ma.hw.ac.uk} \\
Department of Mathematics, Heriot-Watt University \\
Edinburgh EH14 4AS, United Kingdom } \\

\vspace{0.5cm}

{ April   2002}

\end{center}

\begin{abstract}

\noindent Starting with the Chern-Simons formulation of (2+1)-dimensional 
gravity we  
 show that the gravitational interactions deform
the Poincar\'e symmetry of flat space-time to a quantum group 
symmetry. The relevant quantum group is the quantum double 
of the universal cover of the (2+1)-dimensional Lorentz group,
or Lorentz double for short. We construct the
Hilbert space of two gravitating particles 
and  use the universal $R$-matrix of the  Lorentz double to 
derive a general expression for the  scattering cross section of  gravitating 
particles with spin.  In appropriate limits our formula reproduces
the  semi-classical scattering formulae found by 't~Hooft, Deser, 
Jackiw and de Sousa Gerbert.

\end{abstract}


\section{Introduction}
The extensive literature on  (2+1)-dimensional gravity (see \cite{Carlipbook}
for a  bibliography) includes
numerous hints that quantum groups should play a pivotal role in a
satisfactory formulation of the quantised theory. Physically, an
important  hint comes from  the  role of the braid group in the
classical interaction of gravitating point particles in two
spatial dimensions, see e.g. \cite{Carlipscat}. 
Since quantum groups  yield representations of
the braid group, this suggest a natural place for quantum groups
in the quantised theory.  At the mathematical level, the
possibility of formulating (2+1)-dimensional gravity as a
Chern-Simons theory gives a strong clue. The quantisation of
Chern-Simons theory with compact gauge groups is a long  and
beautiful chapter of mathematical physics (not yet closed) in
which quantum groups are central protagonists.

The goal of this paper is to establish the central role
of a certain quantum group  in  (2+1)-dimensional  quantum gravity. 
We  show that the quantum double of the
 Lorentz group in 2+1 dimensions (or Lorentz double for short) 
arises automatically in the quantisation of (2+1)-dimensional
gravity with vanishing cosmological constant
and  demonstrate that it provides a powerful new tool for
analysing interesting physics. In this paper we focus
on the gravitational scattering of particles. We show that 
in the quantum theory gravitating particles  carry representations 
of the Lorentz double and that, as a result, the multi-particle
Hilbert space carries a representation of the braid group.
In a rather precise sense, gravitating particles in two spatial dimensions
 may thus be thought of as  gravitational non-abelian anyons.
We show that the scattering is determined by the representation of 
the braid group.

In some ways this paper is a continuation
of \cite{BM}, where the Lorentz double was first introduced  and
its role in understanding (2+1)-dimensional gravity was pointed
out. Two questions were raised  but not answered in that paper.
The first concerns the role of the Lorentz double in a 
deformation quantisation of the classical gravitational phase
space. That question was addressed  in the Euclidean setting in
\cite{schroers} and here we will be able to adapt the results of that
paper  to
the Lorentzian situation without much difficulty. The second,
physically deeper, question concerns the reconstruction of
space-time physics (such as scattering) 
from the rather abstract  representation theory
of the Lorentz double. This is the main  topic of the current
paper,  which is organised as follows.

We begin in the next section 
by comparing the interactions of charge-flux composites
with those of gravitating particles in 2+1 dimensions, pointing 
out the topological nature of both. The purpose of that section
is to motivate our 
approach to (2+1)-dimensional quantum gravity and to outline,
without much technical detail, some of the gravitational
 phenomena we hope to explain using quantum groups.  The 
technical development begins in Sect. 3. 
 We review the representation theory of the Poincar\'e group 
in  2+1 dimensions and introduce 
the basic notion of ribbon Hopf algebras.  We show how they
provide the natural language for discussing spin and statistics
of particles in 2+1 dimensions. Using and generalising  results 
of \cite{Fredenhagen}, \cite{FM} and \cite{MS} we give 
a relativistic treatment of the general Aharonov-Bohm scattering
of particles which carry representations of ribbon Hopf algebras.
The level of exposition in this section is quite technical and 
detailed  since it provides the basis of our analysis of 
quantised gravitational interactions in 2+1 dimensions. An important
aspect of our treatment is that it takes places entirely in momentum
space.
In sect. 4  we briefly 
review  the Chern-Simons formulation of (2+1)-dimensional gravity
with a particular emphasis on the inclusion of particles. 
Following the approach pioneered by Fock and Rosly \cite{FR}
we  sketch how the  the Poisson 
structure of the phase space of (2+1)-dimensional gravity can be
described  in
terms of a classical $r$-matrix. Our approach to the 
quantisation of 
the phase space is motivated by the combinatorial quantisation programme 
developed in \cite{AGSI}, \cite{AGSII} and \cite{AS} for Chern-Simons
theories with compact gauge groups. An important 
step in that quantisation procedure is the identification of a 
quantum $R$-matrix (solution of the quantum Yang-Baxter equation)
which reduces to the classical $r$-matrix of the Fock-Rosly 
description in the classical limit. In sect. 5 we introduce
the Lorentz double and show that its $R$-matrix has the desired
limit. We do not fully work out the combinatorial quantisation
programme for (2+1)-dimensional gravity, which requires further 
technical analysis. Having identified the Lorentz double as 
a central ingredient in the quantisation we use it instead 
 as a tool for  addressing  the 
physical issues raised in sect. 2. In sect. 6 we construct 
 the theory of gravitationally
interacting particles  in  analogy with the theory of 
relativistic anyons developed in sect. 3.
This analogy is both very powerful and subtle:
ordinary anyons carry representations of both
the Poincar\'e group and a ribbon Hopf algebra, 
but our gravitational
anyons only carry representations of the Lorentz double.
The Lorentz double thus plays a dual role, determining 
both the space-time properties and the braiding of gravitating
particles in 2+1 dimensions. To illustrate and test the
theory we construct the  Hilbert space of two massive particles
with arbitrary spins and compute the scattering cross sections.
In sect. 7 we discuss our results and draw some conclusions. 

Throughout the paper we use units in which the speed of 
light is 1. We mostly   set $\hbar =1$ and in the gravitational
calculations we  use the fact that Newton's constant in 
(2+1)-dimensional gravity has the dimensions of inverse mass
to measure masses in units of $(8\pi G)^{-1}$. The Planck length
in 2+1 dimensions is ${\ell}_P=\hbar G$.
Whenever the interplay of the constants 
$\hbar$ and $G$ in  our calculations seemed interesting 
we have explicitly included  them.

\section{Topological interactions in 2+1 dimensions}

\subsection{An algebraic approach to Aharonov-Bohm scattering}

The best known  manifestation of non-trivial
 topological interactions in a physical experiment 
is the Aharonov-Bohm interaction between an electrically charged
particle and a magnetic flux tube.  More generally, such 
interactions occur between flux-charge composites $(\Phi,q)$.
Here $\Phi\in[0,2\pi)$ labels the fractional part of the magnetic
flux and $q$ is the integer electric charge. The point of the following
brief review is to show how the familiar Aharonov-Bohm interaction
of such composites  can be understood from 
an algebraic perspective. While some of the 
algebraic language may seem unnecessarily complicated in this 
simple example it will help to  highlight those aspects
which generalise to the gravitational interaction.

The starting point of the algebraic interpretation is the 
interpretation of the label $(\Phi,q)$ as the representation label
of the ``double'' 
\bea
D=\ZZ \times U(1).
\eea
For what we are about to say we  should really replace $U(1)$ by
its group algebra, but to keep the discussion as simple as possible
we think of $D$ simply as an abelian group. We write its  elements
as $(n,\omega)$, $n\in \ZZ$ and $\omega\in [0,2\pi)$, and the 
group composition is simply 
\bea
(n_1,\omega_1)(n_2,\omega_2) = (n_1+n_2,\omega_1+\omega_2),
\eea 
where the addition of the angles should be taken modulo $2\pi$.
Irreducible  
representations (irreps) of $D$  are simply tensor products of 
the one dimensional irreps of $\ZZ$ and of $U(1)$. They are 
therefore labelled by a angle  $\Phi\in [0,2\pi)$ and an integer $q$, which
we interpret physically  as  flux and charge. Writing $W_{(\Phi,q)}$
for the carrier space of the irrep $(\Phi,q)$, the action of
 $(n,\omega)\in D$ on $\phi \in W_{(\Phi,q)}$ is 
\bea
r_{(\Phi,q)}(n,\omega)\phi = e^{i(\Phi n + q \omega)} \phi.
\eea
Physically, one of the most basic observed properties 
of flux-charge composites  are the rules for 
 combining flux-charge composites, often called fusion rules.
These   are determined
by the decomposition of the tensor product 
  into irreps,  which 
in turn is ruled by the co-multiplication of $D$. This is 
a map
\bea
\Delta: D\rightarrow D\otimes D,
\eea
allowing elements of $D$ to act 
on $W_{(\Phi_1,q_1)}\otimes W_{(\Phi_2,q_2)}$. 
In the present situation, the co-multiplication is so
simple (often called group-like) that it is not usually mentioned 
explicitly:
\bea
\Delta((n,\omega)) = (n,\omega)\otimes (n,\omega).
\eea
It follows that, at the level of representations,  
\bea
\label{sumrule}
(\Phi_1,q_1)\otimes(\Phi_2,q_2) \simeq
(\Phi_1+\Phi_2,q_1+q_2)
\eea
showing that fluxes and charges simply add when flux-charge
composites are fused.

One of the physically most striking and interesting properties
of charge-flux composites is that they can carry fractional
spin and that they undergo non-trivial Aharonov-Bohm scattering.
To see how this is encoded in the algebraic structure of $D$
we define the element
\bea
c={1\over 2 \pi}\int d \omega\sum_{n\in \ZZ}e^{-i\omega n}(n,\omega).
\eea
(In sect. 5  we will see how to make sense of the infinite sum.)
Acting  with $c$ on $\phi \in W_{(\Phi,q)}$  we find
\bea
r_{(\Phi, q)}(c)\phi = e^{i\Phi q}\phi.
\eea
The eigenvalue $e^{i\Phi q}$ is precisely the ``spin factor''
associated to a flux-charge composite. It determines the fractional
part of the spin $s$ carried by the flux-charge composite  $(\Phi,q)$
according to the rule
\bea
\label{spinrule}
e^{2\pi i s} = e^{i\Phi q}. 
\eea

The final ingredient for discussing statistical properties algebraically
is the algebraic equivalent of the  monodromy operation, which 
transports one flux-charge composite around each  other. This is 
implemented by the element $Q\in D\otimes D$ given by
\bea
Q={1\over (2\pi)^2}\sum_n\sum_{n'}\int d\omega \int d\omega' \,\
e^{-i(\omega n +\omega'n')}(n,\omega')\otimes (n',\omega).
\eea
The element $c$ and the monodromy operation operator $Q$ are linked
by the important  relation
\bea
\label{spinstat1}
\Delta c = Q \,\,c\otimes c.
\eea
This relation, called the generalised spin-statistics relation,
shows that our interpretation of $c$ and $Q$ is consistent:
rotating a system of two flux-charge composites by $2\pi$ is the same 
as rotating the composites individually by  $2\pi$ and 
 transporting  the composites around each other.

Now consider  two flux-charge composites  $(\Phi_1,q_1)$  and $(\Phi_2,q_2)$
with spins $s_1$ and $s_2$  both obeying the condition (\ref{spinrule}).
By (\ref{sumrule})
the total flux and charge of the combined system is $(\Phi,q)=
(\Phi_1+\Phi_2,q_1+q_2)$. It follows from (\ref{spinstat1}) (and also
by direct computation) that 
the action on an element $\phi_1\otimes\phi_2\in 
W_{(\Phi_1,q_1)}\otimes W_{(\Phi_2,q_2)}$
is 
\bea
r_{(\Phi_1, q_1)}\otimes r_{(\Phi_2, q_2)}(
Q)\phi_1\otimes\phi_2 = e^{i(\Phi_1 q_2+ \Phi_2 q_1)}\phi_1\otimes \phi_2.
\eea
If we impose the spin quantisation rule
(\ref{spinrule}) also on the spin $s$ of the combined system
it follows that 
\bea
e^{2\pi i(s-s_1-s_2)}=e^{i(\Phi_1 q_2+ \Phi_2 q_1)}.
\eea
However, $s-s_1-s_2$ is just the orbital angular momentum $l$ of the relative
motion of the two flux-charge composites in the centre of mass frame. 
We therefore conclude 
 that the orbital angular momentum for the relative motion
of flux-charge composites $(\Phi_1,q_1)$ and $(\Phi_2,q_2)$ 
has to satisfy the quantisation 
condition
\bea
e^{2\pi i l}=e^{i(\Phi_1 q_2 + \Phi_2 q_1)},
\eea
which implies
\bea
l=n+{1\over 2 \pi}(\Phi_1 q_2 + \Phi_2 q_1).
\eea
The important consequence of the spin  condition (\ref{spinrule}) and 
the spin-statistics relation (\ref{spinstat1}) is thus that not 
only the individual spins of flux/charge composites 
 but also the orbital angular momentum of their relative motion 
has a fractional part. It is this fractional part which is 
responsible for their non-trivial Aharonov-Bohm scattering.

A careful discussion of the 
phase-shift analysis of the standard Aharonov-Bohm scattering 
problem, where a beam of electrons  of charge $q_1=e$ is scattered
off a tube carrying magnetic flux $\Phi_2=\Phi$ can be found in \cite{Hagen}.
The non-integer eigenvalues  $l=n+\frac{e\Phi}{2\pi}$
of the orbital angular momentum are a basic but crucial input to
the discussion. 
The phase shift in the $n$-the partial wave, or equivalently the restriction
of the $S$-matrix to  the $n$-the partial wave depends on
$\frac{e\Phi}{2\pi}$ and on the sign of $[l]$, the
 largest integer $\geq l$, \cite{Hagen}: 
\bea
\label{ABS}
S^{(l)}
=e^{2i\delta_l}=\left\{ \begin{array}{ll}e^{-{i\over 2}e\Phi}  & \mbox{if}
\quad [l]\geq 0  \\
e^{{i\over 2}e\Phi}  & \mbox{if}\quad
[l] < 0.  \\
\end{array} \right.
\eea
The
resulting normalised cross section in the frame of the flux tube is 
the famous Aharonov-Bohm cross section
\bea
\frac {d\sigma}{d\varphi}(\varphi)=\frac{\hbar }{2\pi M_e v}\frac{\sin^2\left(
\frac{e\Phi}{2}\right)}{\sin^2\frac{\varphi}{2}},
\eea
where $M_e$ and $v$ are the mass and the speed of the incident
electrons
and $\varphi$ is the scattering angle.

\subsection{Gravitational scattering in 2+1 dimensions }

One reason why gravity in 2+1 dimensions is so much simpler than
in 3+1 dimensions is that the Ricci tensor has the same number 
of independent components as the Riemann tensor. Thus the curvature
of space-time is entirely determined, via the Einstein equations, 
by the energy-momentum tensor. 
 The simplest situation which illustrates
this fact, and moreover gives the first hint that what we have 
summarised in the preceding sections is indeed relevant  to (2+1)-dimensional
gravity, is the motion of a test particle in the 
gravitational field of a point particle at rest. 

In a local coordinate system with coordinate indices $a, b =0,1,2$,  and 
in units where the speed of light is 1 
the Einstein equations read
\bea
\label{Einstein}
R_{ab}-{1\over 2}g_{ab} R=8\pi G T_{ab}.
\eea
Here $R_{ab}$ is the  Ricci tensor, $R$ is the Ricci scalar
and $T_{ab}$ is the energy-momentum tensor. As mentioned in the 
introduction,  Newton's constant
$G$ has the dimension of inverse mass  in 2+1
dimensions. For a particle of mass $M $ we can therefore  
define the dimensionless quantity
\bea
   \mu= 8\pi G M, 
\eea
which measures the mass 
 in units of $1/(8\pi G)$.
Suppose  a point particle of mass $M$ is at rest 
 at the origin of a local coordinate system
$(x_0,x_1,x_2)$. Einstein's equation for this situation are
\bea
\label{pointsource}
R_{ab}-{1\over 2}g_{ab}=\mu\delta(x_1)\delta(x_2).
\eea 
They were first solved in 
\cite{Staru}. The space-time surrounding the particle is flat,
but has a conical singularity at the particle's position. The cone's
deficit angle is determined by the mass.  In terms of 
polar coordinates $(\rho,\varphi)\in\RR^+\times[0,2\pi-\mu)$ 
the spatial line element is just 
\bea
ds^2=(d \rho )^2+\rho^2(d\varphi)^2.
\eea
For some  purposes  another coordinate system  $(r,\theta)$ used
in \cite{DJ} is more  useful. It describes the cone  embedded
in flat three-dimensional space in terms of the  its projection
parallel to its axis of symmetry. 
Defining the rescaling parameter
\bea
\label{useful}
\alpha =1-{\mu\over 2\pi},
\eea
the projected   coordinate are $(r,\theta)$, where  $r =\alpha \rho$
and  $\theta=\varphi/\alpha$,
which has the usual angular range $\theta\in [0,2 \pi)$.
The line element now has the form
\bea
ds^2=\alpha^{-2}(dr^2)+ r^2(d\theta)^2.
\eea
Geodesics are most easily visualised by cutting the cone open,
unfolding it and drawing 
straight lines. Upon gluing the cone back together and looking at the 
geodesics in the projected coordinates system $(r,\theta)$ 
one finds that  some of the
geodesics  cross.  One can interpret the result by saying that parallel 
geodesics get deflected by an angle $\pm \mu/2\alpha$, with the 
sign only  depending  on whether
they have passed the cone's apex on the left or on the right
but not on the geodesic's distance from the apex. 
Such distance-independence is typical of topological interactions
and  reminiscent of the phase shifts in 
Aharonov-Bohm scattering.
The analogy can be made more convincing
by considering the corresponding quantum problem. This was first
done in \cite{Hooft} and \cite{DJ}. In both papers the mathematical
model is the Schr\"odinger equation on the cone. For stationary scattering 
states with energy $E=\hbar^2k^2/2 M$ this becomes 
\bea
\left( \left(r {\partial \over \partial r}\right)^2 + 
\left({ 1\over \alpha}{\partial \over \partial \theta}\right)^2 +
\left({k r \over \alpha }\right)^2\right)\psi =0.
\eea
The basic but crucial input in the partial wave analysis of 
the scattering problem in \cite{DJ} is the observation 
that the orbital angular momentum operator 
$-i{ 1\over \alpha}{\partial \over \partial \theta}$ has non-integer
eigenvalues
\bea
\label{lquant}
l={n\over \alpha}.
\eea
This is responsible for  non-trivial  phase shifts and hence a
non-trivial
$S$-matrix:
\bea
\label{DJshifts}
S^{(l)}=e^{2i\delta_l} =
\left\{ \begin{array}{ll}e^{-{i l \mu \over 2 }}  & \mbox{if}
\quad [l] \geq 0  \\
e^{{i l \mu \over 2 }}  & \mbox{if}\quad
 [l] < 0.  \\
\end{array} \right.
\eea
While this formula is reminiscent of the Aharonov-Bohm phase
shifts (\ref{ABS}), there are important differences, most 
notably the increase  of the phase shift with the magnitude of the 
 angular momentum $l$
 in (\ref{DJshifts}). In the rest of this paper we will show that 
the gravitational phase shifts nevertheless have an interpretation
in terms of spin and statistics,  much
like the Aharonov-Bohm phase shifts.
We begin by looking  at  a generalisation of the 
 Aharonov-Bohm effect from a more 
abstract, algebraic perspective.

\section{Spin sum rules and relativistic scattering of particles obeying 
braid statistics}

\subsection{The Hilbert space of a single relativistic anyon}
The Aharonov-Bohm story generalises naturally to 
particles in 2+1 dimensions
carrying representations of ribbon Hopf algebras. 
We are interested in relativistic particles, and since the 
consequences of relativistic invariance are crucial in this paper,
we briefly review them. 

Write Mink$_3$ for three-dimensional Minkowski space  and denote
its elements by  $\bx= (x_0,x_1,x_2)$ and  $\by
=(y_0,y_1,y_2)$. The inner product is 
\bea 
\eta(\bx,\by) =x_0y_0-x_1y_1-x_2y_2. 
\eea 
The  group of linear
transformations on Mink$_3$ leaving $\eta$ invariant is the group
$O(2,1)$ which has four connected components. The Lorentz group 
(sometimes called the proper Lorentz group) is
the  subgroup $SO(2,1)$ of $O(2,1)$ transformations of determinant
one. It still has two components, one which preserves the
direction of time and one which reverses it. The component
containing the identity consists of those $SO(2,1)$
transformations which preserve the direction of time. It is called
the orthochronous Lorentz group and we denote it by $L_3^\uparrow$. The
orthochronous Lorentz group is infinitely connected. Its double cover
$SU(1,1)$ and its universal cover $\tilde L_3^\uparrow$  are
described in detail in appendix A. 
We denote elements of $\tilde L_3^\uparrow$ 
by $u,v,w,x...$ and the  $\Lor$-element associated to 
$u\in\tilde L_3^\uparrow$ by
$\Lambda(u)$. 

The group of affine transformations on Mink$_3$ leaving 
$\eta$ invariant is the  semi-direct product $\RR^3 \rtimes O(2,1)$
of translations and $O(2,1)$ transformations. For us the identity
component of that group is particularly relevant. This is the 
 orthochronous Poincar\'e group
\bea 
P_3^\uparrow = \RR^3 \rtimes L_3^\uparrow, 
\eea 
which we shall often simply refer to as the Poincar\'e group.
The multiplication law for two elements $(\ba_1, \LL_1),
(\ba_2,\LL_2)\in P_3^\uparrow$ is 
\bea 
\label{poincaremulti}
(\ba_1,\LL_1)(\ba_2,\LL_2)=
(\ba_1+\LL_1\ba_2,\LL_1\LL_2). 
\eea 
The universal cover of the orthochronous Poincar\'e group is
\bea
\tilde P_3^\uparrow = \RR^3 \rtimes \tilde L_3^\uparrow.
\eea
For two elements $(\ba_1, u_1),
(\ba_2,u_2)\in \tilde P_3^\uparrow$   the multiplication rule is 
\bea
\label{covermulti}
(\ba_1,u_1)(\ba_2,u_2)=
(\ba_1+\Lambda(u_1)\ba_2,u_1u_2). 
\eea 

Following the notation  used in the
literature on (2+1)-dimensional gravity we denote the Lie algebra
of $L_3^\uparrow$ by $so(2,1)$.
We denote the Lie algebra of $P^\uparrow_3$ 
by $iso(2,1)$ and introduce generators $P_a$, $J_a$,
$a=0,1,2$, satisfying the commutation relations 
\bea
\label{poincarecom}  [P_a,P_b] =0,
\quad[J_a,J_b]=\epsilon_{abc} J^c,\quad 
[J_a,P_b]=\epsilon_{abc} P^c,
\eea 
where
$\epsilon_{abc}$ is the totally antisymmetric tensor in three
dimensions, normalised so that $\epsilon_{012}=1$.
The elements  $P_0$, $P_1$ and $P_2$ 
generate translations in time and space, $J_0$ generates spatial
rotations and $J_1$ and $J_2$ generate boosts.  
In the context of (2+1)-dimensional gravity it is  important
that we can interpret the  Lie algebra $so(2,1)$ as momentum space
by associating to a momentum $\bp=(p^0,p^1,p^2)$ 
the element $p^0J_0+p^1J_1+p^2J_2\in so(2,1)$. The Lorentz group acts
on this momentum space  via the adjoint representation, leaving the 
inner product $\bp^2=\eta(\bp,\bp)$ invariant.
 Thus $J_0$ is also  the generator of rotations in momentum space,  but 
 with our conventions $[J_0,J_1]=-J_2$
so $J_0$  generates a clockwise rotation
in the $p_1p_2$ plane.   $J_1$ is the generator
of boosts along the  $2$-axis (because $[J_1,J_0]=J_2$)
 and  $J_2$ is the generator of 
boosts along the negative $1$-axis (because  $[J_2,J_0]=-J_1$). 
For more details on our conventions we 
refer the reader to appendix A.

According to Wigner's dictum the fundamental properties 
mass and spin of a  particle
in  (2+1)-dimensional Minkowski space label  irreps 
of the universal cover of the (2+1)-dimensional 
Poincar\'e group. The irreps  of  the (2+1)-dimensional 
Poincar\'e group are described in  \cite{BR} and the 
representation theory of  he universal cover is studied in some detail in 
\cite{Grigore}.
Some are not physically relevant, but those which are,  
 are  indeed labelled 
by a positive and real parameter $\MM$, interpreted as the 
particle's mass,
and another real parameter   $s$ interpreted as the particle's 
spin. 
We denote these irreps by 
$\pi_{\MM s}$.
Geometrically, $\MM$ labels an orbit of $\tilde  L_3^\uparrow$
acting on  momentum space $(\RR^*)^3$. This orbit is obtained by
boosting the energy-momentum vector $(\MM,0,0)$ of a particle at  rest,
thus sweeping out
the mass hyperboloid
 \bea
\label{massshell}
H_\MM=\{\bp\in (\RR^*)^3|
 \bp=\LL(\MM,0,0)^t,\,\LL \in  L_3^\uparrow\}.
\eea  
The 
mass hyperboloid consists  of all momenta $\bp$ satisfying $\bp^2
=\MM^2$ and $\bp_0 >0$.
Geometrically, it can be 
identified  with the coset $\Lor/SO(2)=\LLor/\RR$. 
Using the parametrisation (\ref{minkeulerr}) of Lorentz transformations
we can therefore 
write elements of $H_\MM$ in terms of a boost parameter $\vartheta$
and an angle $\varphi$ as 
\bea
\label{momentum}
\bp=(\MM\cosh \vartheta, \MM \sinh\vartheta\cos\varphi,
 \MM \sinh\vartheta\sin\varphi)^t.
\eea

The spin  $s$ labels
an irrep  of the centraliser group
of the reference momentum $(\MM,0,0)$  on the mass hyperboloid.
The centraliser group is  the universal cover of the 
two-dimensional rotation group $\widetilde{SO(2)}\cong\RR$.
Hence the spin $s$ takes arbitrary real values in the representation
theory of the universal cover of the Poincar\'e group. In the 
representation theory of the Poincar\'e group itself the spin is an
integer. The irreps  $\pi_{\MM s}$ 
of $\tilde P_3^\uparrow$  can be described in two equivalent ways.
Both require a Lorentz invariant  measure $dm(\bp)$  on the  mass hyperboloid 
(\ref{massshell}).
Using the parametrisation (\ref{momentum}) we can write that measure
in terms of the  boost parameter $\vartheta\in \RR$ and the
 angle $\varphi \in [0,2\pi)$
as 
\bea
\label{measure}
dm(\bp) =\frac {\MM }{8\pi^2}
\sinh \vartheta d\vartheta d \varphi.
\eea
As we shall see, 
the normalisation of the measure is motivated by the usual 
normalisation of  scattering states in relativistic quantum theory.
The irreps of the universal 
cover of the Poincar\'e group can 
be described either in terms of  equivariant
functions on $\LLor$ or in terms of functions on 
 $H_\MM$ on which Poincar\'e transformation 
act via a multiplier representation \cite{BR}.
For us both points of 
view  are important. 

First we consider equivariant functions and  define
\bea 
\label{Hilbert}
V_{\MM s}=\{ \phi : \tilde  L_3^\uparrow \rightarrow \CC)|\phi(x r(\psi))&=&
e^{-i\psi s} \phi(x) \,\,\,\forall\,\psi \in \RR, x\in \LLor,
\nonumber \\ &&
 \int_{H_\MM}|\phi(x)|^2\,dm(\bp) <\infty \},
\eea 
where $r(\psi)$ is the anti-clockwise (mathematically positive)
spatial rotation by $\psi$ as  defined in 
(\ref{rotations}). Note that integrating $|\phi(x)|^2$ with respect
to $dm(\bp)$ makes sense since it only depends on
 the projection $\bp=\Lambda(x)(\MM,0,0)^t$. 
Elements $(\ba,u)\in \tilde P_3^\uparrow$ act on this space via
\bea
\label{Prep}
\pi_{\MM s}(\ba,u)\phi(x)= \exp(i\ba\cdot\bp)\phi(u^{-1}x).
\eea
where $\bp$ is again he momentum on the mass shell $H_\MM$ 
associated to  $x\in\LLor$ via the projection $\bp=\Lambda(x)(\MM,0,0)^t$.

Alternatively, we can describe irreps in terms of  the Hilbert
space $L^2(H_\MM,dm)$ of square integrable functions on  the 
mass hyperboloid. The action of $\rho_{\MM s}(\ba,u)$ on 
an element  $\Psi \in L^2(H_\MM,dm)$   is 
\bea
\label{Rrep}
\rho_{\MM s}(\ba,u)\Psi(\bp)= \exp(i\ba\cdot\bp)\exp({is\psi(u,\bp)})
\Psi(\Lambda(u^{-1})\bp),
\eea
where the   phase factor  $\exp({is\psi(u,\bp)})$ 
encodes the spin of the representation 
$\rho_{\MM s}$ and is called the multiplier.  To write down an 
explicit formula for the multiplier and to establish an isomorphism
between the representations $\pi_{\MM s}$ and $\rho_{\MM s}$ 
one requires a measurable map
\bea
s: H_\MM\rightarrow \LLor
\eea
satisfying $\Lambda(s(\bp))(\MM,0,0)^t=\bp$, see
\cite{BR} for details.
Geometrically, $s$ is a section
 of the bundle $\LLor \rightarrow H_\MM$.
The phase  $\psi(u,\bp)$  entering the multiplier  is defined  in terms
of $s$:
\bea
\label{multiplierphase}
r(\psi(u,\bp))=(s(\bp)^{-1}u \, s(\Lambda(u^{-1})\bp)). 
\eea

Having provided the general framework for space-time characteristics
of particles we turn to 
the  generalisation of the internal charges $(\Phi,q)$ encountered 
in the Aharonov-Bohm effect. This requires
some Hopf algebra technology.
We refer the reader
to \cite{CP} for a full definition of a ribbon Hopf algebra. 
For our purposes it is sufficient to know that a ribbon Hopf algebra ${\cal A}$
combines and generalises the algebraic elements which entered the 
discussion  of flux/charge composites.  Thus  there is a multiplication
rule 
\bea
m: {\cal A}\otimes {\cal A}\rightarrow {\cal A}
\eea
and a co-multiplication
\bea
\Delta: {\cal A}\rightarrow {\cal A}\otimes {\cal A}
\eea
which dictates the way tensor products of representations are decomposed.
Ribbon Hopf algebras are in particular quasi-triangular Hopf algebras, which
means that there exists an  invertible element 
$R\in {\cal A}\otimes {\cal A}$,
called the universal $R$-matrix, which obeys the quantum Yang-Baxter equation
\bea
\label{qybe}
R_{12}R_{13}R_{23}=R_{23}R_{13}R_{12}.
\eea
Here $R_{12}=R\otimes 1\in {\cal A}\otimes {\cal A}\otimes {\cal A}$ 
and $R_{13}$ and $R_{23}$
are defined similarly. 
Finally, ribbon Hopf algebras have a special central element $c$ 
which together with $R$ obeys the generalised spin-statistics relation
\bea
\label{spinstatrel}
\Delta c=  (R_{21} R) c\otimes c.
\eea
The element $R_{21}\in {\cal A }\otimes{\cal  A}$ is 
obtained from $R$ by changing 
the order of the tensor product. If $ R=\sum_i a_i\otimes b_i$ then
$R_{21}=\sum_i b_i\otimes a_i$. The product $R_{21}R $ is the monodromy
operator. We introduce the abbreviation 
\bea
\label{monodromy}
Q:=R_{21}R. 
\eea

Consider now a particle  in two spatial dimensions
which is charged with respect to a ribbon Hopf algebra ${\cal A}$. By
this we mean that it carries an irrep $r_{A}$
of ${\cal A}$. The  internal state of the particle
is an element of the carrier space $W_A$ 
of $r_A$ and 
the kinematic state  of the particle is an element of 
an  irrep $V_{\MM s}$ of $\tilde P^\uparrow_3$ 
labelled by its mass $\MM$ and its spin $s$. 
We have kept the following discussion  general but restrict attention
to unitary  irreps  and assume that tensor products of
representations are reducible into irreps.  

The 
central  ribbon element $c$ of the internal symmetry 
${\cal A}$  acts on 
$W_A$ via a phase:
\bea
r_A(c) W_A = e^{2\pi i s_A} W_A.
\eea
The Poincar\'e group $\tilde P^\uparrow_3$ 
has a centre isomorphic to $\ZZ$, generated
by the $2\pi$-rotation, which we denote by  $\Omega$. Being central,
$\Omega$, too,  
acts  by a phase:
\bea
\pi_{\MM s}(\Omega)\, V_{\MM s} = e^{2\pi i s}V_{\MM s}.
\eea
The physically allowed  states of the particle are 
determined by demanding 
 the spin quantisation condition
\bea
\label{spinquant}
e^{2\pi i s_A} =e^{2\pi i s}
\eea
i.e. $s=s_A + n$, where $n \in \ZZ$.
There is no {\it a priori} restriction on the value of $s_A$,
showing that particles which carry representations of ribbon Hopf
algebra can have arbitrary spin. In the following we refer to them
as anyons.
Thus, the one-anyon  Hilbert spaces are tensor products of the form
$V_{\MM s}\otimes W_A$ for which $\Omega$ and $c$ give the same phase.
In symbols:
\bea
\label{onehilbert}
{\cal H}_1=
V_{\MM s}\otimes W_A \,\,\,\mbox{provided}\,\,\,
 \pi_{\MM s}(\Omega)\phi\otimes v=
\phi\otimes r_A(c)v \quad \forall \,\, \phi\otimes v\in V_{\MM s}\otimes W_A .
\eea

\subsection{The multi-anyon Hilbert space}

Now consider $n$ anyons, each carrying  an  irrep   
of the internal symmetry  and having   masses $\MM_i$
and spins $s_i$, $i=1,...,n$, all  satisfying the condition (\ref{spinquant}).
The spin sum rules for anyons were analysed in the 
more general framework of algebraic quantum field theory
in  \cite{Fredenhagen},  the multi-anyon Hilbert space was 
constructed for abelian braiding in \cite{FM} and that construction
was extended to  non-abelian
braiding  in \cite{MS} and \cite{FGR}. 
The upshot of these studies is that the multi-particle
Hilbert space is not  isomorphic to the tensor product 
of single anyon Hilbert spaces. Instead the $n$-anyon space
 can be described as the space of 
of $L^2$-sections of a certain vector bundle  over the space of $n$ distinct
velocities. Let $\bq_i=\bp_i/\MM_i$ so that for all $i=1,...,n$
$\bq_i$ lies on the unit mass hyperboloid $\bq_i\in M_1$, and define
\bea
\label{multishell}
{\cal M}_n=\{(\bq_1,...,\bq_n)\in M_1^n|\bq_i\neq\bq_j \quad \mbox{for}
\quad i\neq j\}.
\eea
This is is configuration space of $n$ distinct velocities of $n$
non-identical particles. For identical particles one obtains the 
corresponding configuration space after dividing by the action 
of the symmetric group:
\bea
\label{multishelll}
{\cal N}_n={\cal M}_n/S_n.
\eea
Both spaces are multiply connected. For identical particles
the fundamental group $\pi_1({\cal N}_n)=B_n$ is the braid group on 
$n$ strands, and for non-identical particles the fundamental group 
$\pi_1({\cal M}_n)=PB_n$ is the pure braid group on 
$n$ strands \cite{CP}. Hence there are natural flat $PB_n$ ($B_n$)
bundles over ${\cal M}_n$ (${\cal N}_n$). 
The result of \cite{FM}, \cite{MS} and
\cite{FGR} is that  $n$-anyon Hilbert spaces are the spaces of 
$L^2$-sections of associated vector bundles.
For our purposes it is useful
 to recast  the construction in the  language of ribbon Hopf algebras.
For simplicity we restrict attention to two distinct particles.

The space of velocities  $\bq_1=\bp_1/\MM_1$ and $\bq_2=\bp_2/\MM_2$ 
of two particles
 can be conveniently 
parametrised as follows.
 We  use Lorentz transformations to
go to the rest frame of one of the particles, say particle 1.
In that frame, particle 2 is not generally at rest, but we 
can use rotations to  make sure that the velocity  $\bq_2$ 
is along the positive 1-axis, i.e. $\bq_2$ is of the form
\bea
B(\xi)(1,0,0)^t=(\cosh \xi, \sinh \xi,0)^t,
\eea
where $B(\xi)$ is the Lorentz boost along the $1$-axis with
boost parameter $\xi\in \RR^+$,
i.e.  $B(\xi)=\Lambda(b(\xi))$ in the  notation of (\ref{boosts}). 
The combined momentum  in the rest frame 
of particle 1 is again time-like, so we can write it in the form
\bea
\label{sum1}
(\MM_1+\MM_2\cosh \xi, \MM_2\sinh \xi,0)^t=
V(\xi)(M(\xi),0,0)^t.
\eea
Here  
\bea
\label{totalmass}
M(\xi)=\sqrt{\MM_1^2+\MM_2^2 +2\MM_1\MM_2\cosh \xi}
\eea 
is the 
invariant mass of the combined system.
$V(\xi)$  is a Lorentz boost along the 1-axis
 which relates the rest frame
 of particle  1 to the centre of mass frame. Both depend on $\xi$,
$\MM_1$ and $\MM_2$  but we have suppressed
the dependence on $\MM_1$ and $\MM_2$ in our notation
because both are kept 
fixed in the following discussion. The dependence on $\xi$,
however, is crucial. 
Thus we have an alternative parametrisation of the two-particle
momentum space. Instead of specifying the two momenta $\bp_1$
and $\bp_2$,  we specify the ``relative'' boost parameter $\xi$,
and one overall Lorentz transformation $\LL\in L_3^\uparrow$
which relates $\bp_1$
and $\bp_2$ to their values in the centre of mass frame. 
Explicitly we  define
\bea
\label{relate}
\bq_1(\xi) =V^{-1}(\xi)(1,0,0)^t
\quad \mbox{and} \quad
\bq_2(\xi)=  V^{-1}(\xi)B(\xi)(1,0,0)^t,
\eea
noting  that the two velocities are equal if and only if $\xi=0$.
Thus we have the following bijection
 between the
two parametrisations of the two-particle momentum space:
\bea
\label{bi}
I: L^\uparrow_3 \times \RR^+ &\rightarrow& {\cal M}_2 \\
  (\LL,\xi)&\mapsto& (\LL\bq_1(\xi),\LL\bq_2(\xi)).
\eea

The space ${\cal M}_2$ has a natural Lorentz-invariant measure
which is important for us. It  is the product of the 
usual Lorentz invariant measures on the mass shell of the two 
momenta $\bp_1$ and $\bp_2$.
If we parametrise
\bea
\label{momenta}
\bq_1&=&(\cosh\vartheta_1, \sinh\vartheta_1\cos\varphi_1,  
\sinh\vartheta_1\sin\varphi_1)^t \nonumber \\
\bq_2&=&(\cosh\vartheta_2, \sinh\vartheta_2\cos\varphi_2,  
\sinh\vartheta_2\sin\varphi_2)^t
\eea
then the measure is
\bea
\label{measure2}
dm(\bp_1,\bp_2) = \frac{\MM_1\MM_2}{(8\pi)^2} \sinh\vartheta_1\sinh\vartheta_2
d\vartheta_1d\vartheta_2 d\varphi_1 d\varphi_2. 
\eea
For later use it is important to express this measure also in
terms of the parameters $(\LL,\xi)$ (\ref{bi}). Recall that these
parameters allow us to think of the space ${\cal M}_2$ as foliated
by $\Lor$ orbits, with each orbit parametrised by  a positive boost
parameter $\xi$. Since 
(\ref{measure2}) is invariant under the (left) action of $\Lor$
it is not surprising that the measure is proportional to the standard
invariant measure of $\Lor$. Parametrising 
\bea
\label{com}
\LL =e^{-\theta J_0}e^{-\vartheta J_2}e^{-\varphi J_0}
\eea
one finds the following  remarkably simple   formula
\bea
\label{measure3}
dm(\bp_1,\bp_2) = \frac{\MM_1\MM_2}{(8\pi)^2}\sinh\xi \sinh\vartheta
d\xi d\theta d\vartheta   d\varphi.
\eea
The measure also has a simple expression in terms of the 
total three momentum $\bP=\bp_1+\bp_2$, which can be expressed
in terms of the parameters (\ref{com}) as 
\bea
\label{comformula}
\bP=(M(\xi)\cosh\vartheta,M(\xi)\sinh\vartheta\cos\theta,
M(\xi)\sinh\vartheta\sin \theta),
\eea
 with $M(\xi)$ given by (\ref{totalmass}). In terms of $\bP$
the measure is simply
\bea
\label{measure4}
dm(\bp_1,\bp_2) =\frac{1}{(8\pi)^2 M(\xi)}dP_0\,dP_1\,dP_2\, d\varphi.
\eea

We now have all the ingredients to describe the Hilbert space 
of two distinct relativistic anyons. 
It follows from refs.  \cite{FM} and \cite{MS}
that this Hilbert space can be described   as the space of 
$L^2$-sections of a certain vector bundle over ${\cal M}_2$
associated to the principal bundle 
\bea
\label{pbundle}
\begin{array}{ccc}
PB_2 &\rightarrow & \widetilde{{\cal M}_2} \\
&&\downarrow\\
&&{\cal M}_2.
\end{array}
\eea
In order to  describe this bundle and its section explicitly 
 we need to lift the bijection (\ref{bi})  to 
the universal covers $\widetilde{{\cal M}_2}$ and $\LLor$.
To do this,  first note that $\tilde L_3^\uparrow $ acts on ${\cal M}_2$ and
 that, since $\tilde L_3^\uparrow $ is simply connected, 
 this action lifts to a unique action on $\widetilde{{\cal M}_2}$. 
We denote  elements of $\widetilde{{\cal M}_2}$ 
by $\tilde \bq$  and the action of $ u\in \LLor$ on $\tilde \bq$ 
simply by $u\tilde \bq$. For given $\xi\in \RR^+$
let   $\tilde \bq(\xi)$ be a lift of the reference velocities
$(\bq_1(\xi),\bq_2(\xi))\in {\cal M}_2$
defined in (\ref{relate}).
Then define the lifted bijection via
\bea
\label{bilift}
\tilde I: \LLor \times \RR^+ &\rightarrow&\widetilde{ {\cal M}_2} \\
  (u,\xi)&\mapsto& u\tilde \bq(\xi).
\eea

Consider now the case of two anyons which carry representations $W_A$ and
$W_B$ of the ribbon Hopf algebra $\cal A$. Then
the fibre of the  vector
bundle describing the quantum theory
is $W_A\otimes W_B$. A natural choice of generator for $PB_2$
is the closed path generated by a $2\pi$ rotation in ${\cal M}_2$. 
We denote this generator by $\Omega$. Then we define a representation
of $PB_2$ on $W_A\otimes W_B$ by representing $\Omega^{-1}$ by
$r_A\otimes r_B(Q)$, where $Q$ is 
the monodromy element defined in (\ref{monodromy}).
This representation  defines a vector bundle with fibre
$W_A\otimes W_B$ associated to (\ref{pbundle}).
Sections of this vector bundle are conveniently described 
 in terms of equivariant $W_A\otimes W_B$-valued 
functions  on the cover $\widetilde{{\cal M}_2}$. The equivariance condition
is the requirement that two particle wave functions transform
via $r_A\otimes r_B(Q)$ under the action of $\Omega^{-1}$ on
$\widetilde{{\cal M}_2}$
The two-anyon Hilbert space is thus
\bea
{\cal H}_2=\{\Psi\in L^2(\widetilde{{\cal M}_2},W_A\otimes W_B)|
\Psi(\Omega^{-1}\tilde\bq)=r_A\otimes r_B(Q)\Psi(\tilde\bq\},
\eea
where the  measure on $\widetilde{{\cal M}}$ is again given by the formula
(\ref{measure2}) except that the range of $\varphi$ is now all of $\RR$.

In the case of two particles,
the  description of the Hilbert space can be simplified 
because the structure group $PB_2$ of the bundle is abelian.
As a result the bundle splits into a direct sum of line bundles.
We can perform this splitting explicitly by using the  
decomposition of $W_A \otimes W_B$ into irreps in terms of 
fusion coefficients 
$N_{AB}^C$:
\bea
W_A \otimes W_B = \bigoplus_C N_{AB}^C W_C.
\eea
It follows from the relation (\ref{spinstatrel}) that this decomposition
of $W_A \otimes W_B $ diagonalises the monodromy operator $Q$. For 
each of the fusion channels $W_C$, $N_{AB}^C\neq 0$, the eigenvalue
of $Q$ is given by the combination of spin factors 
$e^{2\pi i (s_C-s_A-s_B)}$. Thus we arrive at the alternative description
of the two-particle Hilbert space
\bea
\label{almostthere}
{\cal H}_2=\bigoplus_C N_{AB}^C\{\Psi\in L^2(\widetilde{{\cal M}_2},W_C)|
\Psi(\Omega^{-1}\tilde\bq)=e^{2\pi i (s_C-s_A-s_B)}\Psi(\tilde \bq)\}.
\eea

We can simplify our description further   by studying the action
of the universal cover of the Poincar\'e group on this space.
 The action of $(\ba,u)\in \tilde P_3^\uparrow$
on $\Psi\in L^2(\widetilde{{\cal M}_2},W_C)$  is
\bea
\rho(\ba,u)\Psi (\tilde \bq)=\exp(i\ba\cdot(\bp_1+\bp_2))
\exp({i(s_1\psi(u,\bp_1)+s_2\psi(u,\bp_2))})
\Psi (u^{-1}\tilde \bq).
\eea
In particular it follows that 
\bea
\rho(\Omega)\Psi (\tilde \bq)= e^{2\pi i(s_1+s_2)} \Psi(\Omega^{-1}\tilde\bq).
\eea
However, since both anyons satisfy the spin quantisation rule (\ref{spinquant})
we can use this relation to rewrite the definition  (\ref{almostthere})
as 
\bea
\label{there}
{\cal H}_2=\bigoplus_C N_{AB}^C\{\Psi\in L^2(\widetilde{{\cal M}_2},W_C)|
\rho(\Omega)\Psi=e^{2\pi i s_C}\Psi\}.
\eea

To cast this definition into its final, most useful form we re-write it
once more using the classical coordinate transformation (\ref{bilift}).
The idea is to use  the pull back
\bea
\tilde I^*:L^2(\widetilde{{\cal M}_2},W_C)\rightarrow 
L^2(\LLor\times \RR^+,W_C),
\eea
which is an isometry of Hilbert spaces. Moreover, defining 
a representation  of $\tilde P_3^\uparrow $ on $
 L^2(\tilde L_3^\uparrow\times \RR^+,W_C)$ via
\bea
\pi(\ba,u)\Phi(x,\xi)=\exp(i\ba\cdot \bP)\Phi(u^{-1}x,\xi),
\eea
where $\bP = \Lambda(x)(M(\xi),0,0)^t$  is the total momentum 
vector (\ref{comformula}),   one checks
that the isometry 
$\tilde I^*$ intertwines between the representations $\pi $  
and $\rho$ of 
$\tilde P_3^\uparrow$.
In particular  we can therefore write the definition (\ref{there})
as 
\bea
\label{final}
{\cal H}_2=\bigoplus_C N_{AB}^C\{\Phi\in L^2(\LLor\times \RR^+,W_C)|
\pi(\Omega)\Phi=e^{2\pi i s_C}\Phi\}.
\eea
This description of the two-anyon Hilbert space has a very
simple interpretation. It is a direct sum of spaces of the form
\bea
V_{s_C}=\{\Phi\in L^2(\tilde L_3^\uparrow\times \RR^+,\CC)\otimes W_C|
\Phi(\Omega^{-1} x,\xi)\Phi=e^{2\pi i s_C }\Phi(x,\xi)\}, 
\eea
each of which
satisfy the equality
\bea
\label{spinsumrule}
\pi(\Omega)V_{s_C} = r_C(c)V_{s_C}.
\eea
This is just the condition we imposed on 
  one-anyon Hilbert spaces (\ref{onehilbert}). 

The spaces $V_{s_C}$ are  reducible as a representation of 
$\tilde P_3^\uparrow$. Its irreducible components are labelled 
by the invariant mass $M$ and spins $s=s_C +n$,
where $n\in \ZZ$.
Thus we have the decomposition
\bea
V_{s_C}=\int_{\MM_1+\MM_2}^{\infty} dM 
\bigoplus_{s=s_C+\ZZ} V_{M s}\otimes W_C.
\eea
into one-particle Hilbert spaces.
Repeating this decomposition 
for every fusion channel $C $ with $N_{AB}^C\neq 0$  we finally 
obtain the decomposition of  
the two-particle space into irreps of $\tilde P_3^\uparrow$ 
and of the ribbon Hopf algebra
${\cal A}$,
each obeying the spin rule (\ref{spinsumrule}):
\bea
\label{twoplektons}
{\cal H}_2 = \bigoplus_C
N_{A B}^C \int_{\MM_1+\MM_2}^{\infty} dM 
\bigoplus_{s=s_C+\ZZ} V_{M s}\otimes W_C.
\eea
In each of the spaces $ V_{M s}$ the quantity
\bea
\label{angmomdef}
l=s-s_1-s_2
\eea
is interpreted as the  orbital angular momentum  of the 
relative motion of the two particles in the centre of mass frame.
Its fractional part plays a crucial role in the scattering theory.

\subsection{Relativistic anyon scattering}

It follows from the 
quantisation condition
\bea
\label{squant}
s=n+s_c, \qquad n\in \ZZ,
\eea
that the orbital angular momentum  of the 
relative motion in the fusion channel $C$ has the
form 
\bea
\label{llquantt}
l=n+\Delta_{AB}^C, \qquad  n\in\ZZ, 
\eea
 where 
\bea
\label{fractional}
\Delta_{AB}^C=s_C-s_A-s_B.
\eea 
Note that $\Delta_{AB}^C$ need not lie in $[0,1)$ so that $[l]$
(the largest integer $\leq l$)
need not be the same as $n$ in the parametrisation (\ref{llquantt}).
Following an argument of E. Verlinde given in a non-relativistic
context in \cite{EVerlinde} we shall now show how to compute
the relativistic scattering of anyons in terms of the numbers
$\Delta_{AB}^C$.   
We  postulate that the $S$-matrix is diagonal in each of the 
components 
$V_{Ms}\otimes W_C$ of the two-particle Hilbert space ${\cal H}_2$.
For given $A,B$ and $C$, $N_{AB}^C\neq 0$,
 its value depends on the sign of 
$l$ and is 
\bea
\label{plekscat}
S^{(l)} =\left\{ \begin{array}{ll}&e^{-\frac i 2\Delta_{AB}^C} \quad
    \,\,\,
\mbox{if}
\quad [l] \geq 0  \\
 &e^{\frac i 2 \Delta_{AB}^C}\qquad  \mbox{if}\quad
[l] < 0.  \\
\end{array} \right.
\eea

In preparation for the discussion of gravitational scattering 
we combine this formula with the relevant phase space factors
to write down  a relativistic cross section for anyon scattering.
We adopt  the conventions and definitions used in 
standard textbooks on quantum field theory such as \cite{IZ}.

As always in scattering theory, we will need to consider
generalised energy and momentum eigenstates which are 
not normalisable and therefore not strictly in  the Hilbert
spaces defined so far. We use standard ket notation for
momentum eigenstates $|\bp\rangle$ of a particle of mass $\MM$.
They can be realised as 
delta-functions on the mass shell (\ref{massshell}) with measure
(\ref{measure}):
\bea
|\bp\rangle=\frac{8\pi^2}{\MM}\frac{1}{ \sinh \vartheta}\delta_{\vartheta}
\,\,\delta_\varphi,
\eea
where $(\vartheta,\varphi)$ are the boost parameter and angle
parametrising $\bp$ via (\ref{momentum}).
These states are normalised so that the inner product of 
two momentum states is 
\bea
\langle\bp'|\bp\rangle=\frac{8\pi^2}{\MM}\,\frac{1}{ \sinh
  \vartheta}\delta_{\vartheta}\,\,
(\vartheta')\delta_\varphi(\varphi').
\eea
Similarly we write two particle scattering states as 
a product of delta functions on ${\cal M}_2$ with measure
(\ref{measure2}):
\bea
\label{twoscatt}
|\bp_1,\bp_2\rangle=\frac{(8\pi^2)^2}{\MM_1\MM_2}
\frac{1} {\sinh \vartheta_1}\delta_{\vartheta_1}
\frac{1} {\sinh \vartheta_2}\delta_{\vartheta_2}
\delta_{\varphi_1}\delta_{\varphi_2}.
\eea
As in the discussion of classical momenta it is convenient
to switch to the parametrisation $(\LL,\xi)$ (\ref{bi})
 in terms of an overall Lorentz transformation and a  relative 
boost parameter. A calculation analogous to the conversion of the
measure into the form (\ref{measure3}) leads to the simple formula
\bea
|\bp_1,\bp_2\rangle&=&\frac{(8\pi^2)^2}{\MM_1\MM_2}\delta_\LL
\frac{1}{\sinh \xi}\delta_\xi \nonumber \\
&=&\frac{(8\pi^2)^2}{\MM_1\MM_2}\frac{1} {\sinh \vartheta}\delta_\vartheta\,
\delta_\theta\,\delta\phi \,\,\frac{1}{\sinh \xi}\delta_\xi, 
\eea
where we have used the parametrisation of $\LL$ given in 
(\ref{com}). Finally note that in terms of the 
 total momentum $\bP=\bp_1 +
\bp_2$  (\ref{comformula}) we have 
\bea
\label{totalmom}
|\bp_1,\bp_2\rangle=(8\pi^2)^2 M(\xi)\delta_{P_0}
\delta_{P_1}\delta_{P_2}\delta_\varphi.
\eea

In order to characterise scattering states in 
the Hilbert space (\ref{twoplektons}) we  need to specify both
the momenta of the anyons and the internal state $v\in W_A\otimes
W_B$. Generic scattering states  of two particles with
momenta $\bp_1$ and $\bp_2$  and internal state $v$ 
are thus  of the form
\bea
|\bp_1,\bp_2;v\rangle =|\bp_1,\bp_2\rangle \otimes v.
\eea  
To simplify our presentation we  consider the case 
where $v$ lies entirely in one of the fusion channels $W_C$ with
$N_{AB}^C\neq 0$.  This assumption and the 
fact that scattering states have a definite invariant
mass $M(\xi)$ means that they are elements of 
\bea
\bigoplus_{s=s_C+\ZZ} V_{Ms}\otimes W_C.
\eea
Looking at the formula  (\ref{totalmom})
and introducing the abbreviation $|s\rangle$ for the function
$e^{-i s\psi}$ on $\LLor$ in the parametrisation
(\ref{minkeuler}) we have the expansion
\bea
|\bp_1,\bp_2;v\rangle & =& 4 (2\pi)^3M(\xi)\delta_{P_0}
\delta_{P_1}\delta_{P_2}\sum_{s=s_C+\ZZ}e^{is\varphi}
|s\rangle\otimes|v\rangle.
\eea
 This expansion is useful
because,  according to our postulate (\ref{plekscat}), 
 the $S$-matrix only depends on the values of $s_C,s_A$
and $s_B$ and no other details of the scattering state.
With the usual decomposition
\bea
\label{sdecomp}
S=1+iT
\eea
we define on-shell  reduced  matrix elements
with respect to an initial  state
$|\bp_1^i,\bp_2^i;v^i\rangle$ and a  final state $|\bp_1^f,\bp_2^f;v^f\rangle$
in the standard fashion
\bea
\label{tred}
\langle\bp_1^f,\bp_2^f;v^f|T|\bp_1^i,\bp_2^i;v^i\rangle=
(2\pi)^3\delta^3(\bP^i -\bP^f)
\langle\bp_1^f,\bp_2^f;v^f|{\cal T}|\bp_1^i,\bp_2^i;v^i\rangle,
\eea
where 
$\bP^i=\bp_1^i+\bp_2^i$ and $\bP^f=\bp_1^f+\bp_2^f$. 
The differential cross section in the centre of mass frame is 
then given by the usual expression \cite{IZ}
\bea
d\sigma=\frac{1}{4\sqrt{(\bp_1\cd \bp_2)^2-\MM_1^2\MM_2^2}}
\int dm(\bp_1^f,\bp_2^f)&\!\!\!(2\pi)^3 &\!\!\!\!
\delta^3(\bP^i -\bP^f) \nonumber
\\
&&
|\langle\bp_1^f,\bp_2^f;v^f|{\cal T}|\bp_1^i,\bp_2^i;v^i\rangle|^2.
\eea
With our simple expression of the $S$-matrix (\ref{plekscat})
we find
\bea
\langle\bp_1^f,\bp_2^f;v^f|{\cal T}|\bp_1^i,\bp_2^i;v^i\rangle
=4 M(\xi) t(\varphi^i -\varphi^f)\langle v^f,v^i\rangle,
\eea
where
\bea
\label{univscat}
i t(\varphi)=\sum_{n}(S^{(l)}-1)e^{is\varphi},
\eea
with $l$ and $s$  expressed in terms of $n$ via (\ref{squant}) and 
(\ref{llquantt}). Explicitly this sum is
\bea
\label{tofphi}
i t(\varphi)= \sum_{n\geq -[\Delta_{AB}^C]}
\left( e^{-\frac i 2\Delta_{AB}^C} -1\right)e^{is\varphi}
+\sum_{n < -[\Delta_{AB}^C]}
\left( e^{\frac i 2\Delta_{AB}^C}-1\right)e^{is\varphi}.
\eea
 This expression   is 
divergent but can be renormalised in  standard fashion \cite{Hagen}.
The resulting finite part  is 
\bea
i \tilde t(\varphi)=e^{-is_C\varphi} e^{-i[\Delta_{AB}^C]\varphi}
\frac{e^{-\frac i 2\Delta_{AB}^C}-e^{\frac i 2\Delta_{AB}^C}}
{1-e^{i\varphi}}.
\eea
This quantity contains  most of the physics of
 the scattering process. The angle $\varphi= \varphi^i-\varphi^f$
is the scattering angle in the centre of mass frame.
It remains to combine the $\varphi$-dependence  with 
appropriate phase space factors to obtain an expression for
the differential cross section. 

Using the expression (\ref{measure4}) for the integration measure
and the formula
\bea
\sqrt{(\bp_1\cd \bp_2)^2-\MM_1^2\MM_2^2}=\MM_1\MM_2\sinh\xi
\eea
we compute
 the following simple formula for the differential 
cross section 
in the centre of mass frame
\bea
\label{hey}
\frac{d\sigma}{d\varphi}(\varphi,\xi)
=\frac{\hbar M(\xi)}{2\pi\MM_1\MM_2\sinh\xi }|t(\varphi)|^2 
|\langle v^f,v^i\rangle|^2,
\eea
where we have reinstated Planck's constant.

Note that in the case where the ribbon Hopf algebra ${\cal A }$ is
abelian and all its irreps one dimensional
the formula simplifies to 
\bea
\frac{d\sigma}{d\varphi}(\varphi,\xi)
=\frac{\hbar M(\xi)}{2\pi\MM_1\MM_2\sinh\xi}\frac {
\sin^2(\frac 1 2 \Delta_{AB}^C)}{\sin^2\frac{\varphi}{2}}.
\eea
The standard Aharonov-Bohm scattering discussed in sect. 2.1 is a
special case of this formula, obtained by setting $s_A=s_B=0$
and $s_C=e\Phi$. 
In order to recover the familiar non-relativistic formula for 
Aharonov-Bohm scattering of electrons off a flux tube  take the limit
where the particles with mass $\MM_1$ (the fluxes) become very heavy
and their rest frame becomes the centre of mass frame. Then
$\MM_1\approx M(\xi)$, $M_2=M_e$  and $\sinh \xi\approx v_2$, the velocity
of the electrons.

\section{The Chern-Simons formulation of gravity in 2+1 dimensions}

\subsection{Einstein gravity as  Chern-Simons theory}
The possibility of writing general relativity
 in  2+1  dimensions as a Chern-Simons
theory was first noticed in \cite{AT}.
This observation opened up a new
approach to gravity and in particular to its quantisation,
which was first systematically explored in \cite{Witten1}.
Since then, a vast
body of literature has been devoted to the subject.
Here we give a brief summary of those aspects which are relevant 
for us.

In  (2+1)-dimensional gravity, space-time is a three-dimensional
manifold $M$.  In the following we shall only consider
 space times of the form $M=\RR\times \Sigma$, where $\Sigma$
is an orientable two-dimensional manifold.
A three-manifold  of that form
is  orientable  and hence, by a classic theorem of  Stiefel,
  parallelisable. Thus its tangent bundle is topologically
trivial.

 In Einstein's original formulation of general
relativity,  the dynamical variable is a  metric $g$ on $M$. For
the Chern-Simons formulation it is essential  to adopt  Cartan's
point of view, where the theory is formulated in terms of the
dreibein of   one-forms $e_a$, $a=0,1,2$ and the spin connection
one-forms $\omega_a$, $a=0,1,2$. The dreibein is related to the
metric via
\bea
\eta^{ab} e_a\otimes e_b \, = g,
\eea
where
$\eta_{ab}=$diag$(1,-1,-1)$. Indices are raised and lowered with
$\eta_{ab}$ and the Einstein summation convention is used
throughout. The  one-forms $\omega_a$ should be thought of as
components of the $L_3^\uparrow$ connection \bea \omega = \omega_a
J^a, \eea where $J_a$, $a=0,1,2$ are the generators  of the Lie
algebra $so(2,1)$ and  satisfy the commutation relations (\ref{poincarecom}). 
 The curvature
two-form $F_\omega = d\omega + \frac{1}{2}[\omega,\omega]$ can be
expanded as $F_\omega= F^a_\omega J_a$, with 
\bea 
\label{spincurv}
F_\omega^a = d\omega^a  + \frac{1}{2} \epsilon^a_{\,bc}
\omega^b\wedge \omega^c. 
\eea 
The Einstein-Hilbert action in
2+1 dimension can be written as 
\bea 
\label{EHaction}
S_{EH}[\omega, e]= \int_M \,\, e_a \wedge F_\omega^a. 
\eea 
In Cartan's formulation of gravity, both the connection $\omega_a$ and the
dreibein $e_a$ are  dynamical variables and
varied independently. Variation with respect to the spin
connection yields the requirement that  torsion vanishes
\bea
\label{notorsion} D_\omega e_a = de_a + \frac{1}{2} \epsilon_{abc}
\omega^b e^c =0. 
\eea
 Variation with respect to $e_a$
yields the vanishing of the curvature tensor
\bea
\label{Einsteineq}
F_{\omega}=0.
\eea
In   2+1 dimensions this is equivalent to the vanishing of the
Ricci tensor, and thus to the Einstein equations in  the
absence of matter.

An important step in the Chern-Simons formulation of gravity
is the  combination of  the dreibein and the spin connection into a
 Cartan connection \cite{Sharpe}. This  is  a  one-form
with values in the Lie algebra $iso(2,1)$ for which 
we defined generators in (\ref{poincarecom}).  
The Cartan connection one-form is
\bea 
\label{Cartan} A = \omega_a
J^a + e_a P^a. 
\eea 
with  curvature
 \bea
\label{decomp} F = (D_\omega e^a) P_a + (F_\omega^a) J_a 
\eea
combining the curvature and the torsion of the spin connection.

The final technical ingredient  we need  in order to establish
the Chern-Simons
formulation is  special to  three dimensional space-times.
This is a non-degenerate, invariant
  bilinear form on the Lie algebra $iso(2,1)$
\bea
\label{inprod}
\langle J_a, P_a\rangle = \eta_{ab}, \quad  \langle J_a, J_b\rangle
= \langle P_a,P_b\rangle = 0.
\eea
Then the  Chern-Simons action  for the
connection
$A$ on  $M=\Sigma\times \RR$ is
\bea
\label{CSaction}
S_{CS}[A] =\frac{1}{2} \int_M \langle A\wedge  dA\rangle
 +\frac{2}{3}\langle A \wedge A \wedge A\rangle.
\eea
A short calculation shows that this is equal to the Einstein-Hilbert
 action (\ref{EHaction}). Moreover, the equation of motion found
by varying the action with respect to $A$ is
\bea
\label{flat}
F=0.
\eea
Using the decomposition (\ref{decomp}) we thus reproduce
the condition of vanishing torsion and the  three-dimensional
Einstein equations, as required.

So far we have only studied the Einstein equations in vacuum.
Matter in the form of point particles
can be included in 
a mathematically elegant fashion in the Chern-Simons
formulation. We refer the reader
to \cite{Carlipscat} for a discussion and  
to  \cite{BM} for further geometrical background.
 Particles are introduced by marking points
on the surface $\Sigma$ and coupling  the particle's phase space
to the phase space of the theory. The phase space of a
 particle with  mass $\MM$ and  spin $s$ is
the set of   energy-momentum vectors $\bp$
 and   generalised angular momenta $\bj=(j^0,j^1,j^2)$ 
satisfying  the mass-shell condition
$\bp^2=\MM^2$ and the spin condition $\bp\cd \bj=\MM s$.
Mathematically, this condition defines  a co-adjoint orbit
  ${\mathcal O}_{\MM s}$ of $P_3^\uparrow$. 
To make this explicit  we  write $P_a^*$ and
$J_a^*$ for the basis elements of
$iso(2,1)^*$ dual to $P_a$ and $J_a$,
and we write an element  $\xi^*\in iso(2,1)^*$ as
\bea
\label{dualphase}
\xi^*=p^a P_a^* + j^a J_a^*.
\eea
Using the inner product
 (\ref{inprod}) we can identify $\xi^*$ with the element
\bea
\label{phaseco}
\xi=p^a J_a + j^a P_a
\eea
in $iso(2,1)$.
With this identification we can think of the 
orbit ${\mathcal O}_{\MM s}$ as lying in $iso(2,1)$.
A generic element like (\ref{phaseco}) can be  obtained by conjugating
the representative element
\bea
\hat \xi=\MM J_0 + s P_0
\eea
with a suitable $P^\uparrow_3$ element $(\ba,L)$.
The orbit  ${\mathcal O}_{\MM s}$ for
 $\MM\neq 0$
 is diffeomorphic to the  tangent bundle of a hyperboloid. 
 Co-adjoint orbits have a canonical
symplectic structure, often called the Kostant-Kirillov
 symplectic structure.
For the generic case $\MM\neq 0$ the corresponding Poisson brackets
of the coordinate functions $j_a$ and $p_a$ are
\bea
\label{kiri}
\{j_a,j_b\}=\epsilon_{abc}j^c, \qquad \{j_a,p_b\}=\epsilon_{abc}p^c.
\eea

In order to introduce $m$ particles with masses and
spins $(\MM_1,s_1),...(\MM_m,s_m)$ we  mark $m$ points $z_1, ...,z_m$
on $\Sigma$ and associate to each point $z_i$
a co-adjoint orbit ${\mathcal O}_{\MM_i s_i}$,$i=1,...m$.
The coupling
of the particle degrees of freedom to the gauge field via minimal
coupling is described in \cite{BM}. The upshot is that
 we specify the kinematic state of each particle in terms of 
representative  $iso(2,1)$ elements
$\hat \xi_{(i)}=\MM_i  J_0 + s_i P_0 $ and Poincar\'e elements
$(\ba_i,\LL_i)$. These determine the curvature of
according to
\bea
\label{gravwith}
F(z) +\sum_{i=1}^m (\ba_i,\LL_i)(8\pi G \MM_i J_0+ s
P_0)(\ba_i,\LL_i)^{-1}\delta^2(z-z_i)=0.
\eea
Expanding the curvature term as in (\ref{decomp}) we find that
the energy-momentum vectors of the particles act
 as  sources for  curvature and their
generalised angular momenta
act as  sources of torsion, in agreement  with physical
expectations. Note that Newton's constant appears as a coupling
constant in such a way that the curvature only sees the rescaled,
 dimensionless masses $\mu_i=8\pi G\MM_i$. The following discussion
is conducted almost entirely in terms of these.

\subsection {Gravitational phase space and its  Poisson  structure}

The phase space of a classical field theory is the space of
solutions of the equations of motions, modulo gauge invariance.
Adopting the  Chern-Simons formulation of  (2+1)-dimensional gravity
we thus find that the phase space of gravity in  2+1 dimensions
is the moduli space of flat $P^\uparrow_3$-connection
on the surface $\Sigma$.
Starting with the classic paper of Atiyah and Bott
\cite{AB}, the  moduli space of flat $G$-bundles on $\Sigma$
 has been studied extensively for
  semi-simple, compact  Lie groups $G$. A pedagogical description of
 that space and its symplectic structure  can be found in \cite{Atiyah}.
The rigorous extension of the theory to non-compact groups like $P^\uparrow_3$
is a very important but (it seems)
 largely open problem.
In the following we give  a brief   description of the
gravitational phase in the   language of moduli spaces of flat connections.
We have endeavoured to be 
conceptually clear but do not enter into technical questions related
to the non-compactness of  $P^\uparrow_3$.

In this paper we are only  interested in the
gravitational interactions of particles, without the
added complication of handles. We therefore
specialise to $\Sigma = \RR^2$ with
$m$ marked points $z_1, ...z_m$ which are not allowed to coincide,
i.e. $z_i\neq z_j$ for $i\neq j$. 
For us, the most useful description of the moduli space is in
 terms of representations of the fundamental group of 
$\RR^2 - \{z_1,...z_m\}$ with a base point $*$ which we choose to 
be at infinity. 
The fundamental group  $\pi_1(\RR^2-\{z_1,...z_m\},*)$ is the 
group freely  generated by $m$
invertible generators $l_1,...l_m$.
A flat $P^\uparrow_3$-connection on $\RR^2-\{z_1,...z_m\}$
associates to each generator a holonomy element in $P^\uparrow_3$. However,
the insertion of the  charges
at marked points as in (\ref{gravwith})
fixes  the holonomy around the $i$-th marked point to be 
 in the $P^\uparrow_3$ conjugacy class
\bea
{\cal C}_i= \{(\ba,\LL)\exp\left(-\mu_i J_0 - s_i P_0\right)
(\ba,\LL)^{-1}|(\ba,L)\in P^\uparrow_3\}.
\eea
Note that for a generic particle with mass $\mu$ and spin $s$, 
we can write elements of the associated conjugacy class as 
\bea
(\ba,\LL)\exp\left(-\mu J_0 - s P_0\right)
(\ba,\LL)^{-1}=\exp(-p^aJ_a-j^aP_a)
\eea
and have the explict formula
\bea
\label{firstmom}
\bp=(\mu\cosh\vartheta,\mu\sinh\vartheta\cos\varphi,
\mu\sinh\vartheta\sin\varphi)
\eea
for the momentum in terms of the parametrisation (\ref{minkeulerr}) of
the Lorentz transformation $L$. The  formula for $\bj$ is 
\bea
\bj=\ba\times\bp +\frac{s}{\mu}\,\,\bp.
\eea

 Fock and Rosly gave a very explicit
description of the Poisson structure of the moduli space 
of flat $G$-bundles on a Riemann surface  in \cite{FR}.
The application of the Fock-Rosly description to Euclidean
 gravity is discussed  in some detail in \cite{schroers}.
It is not difficult to adapt that description to the Lorentzian situation.
To write down the Fock-Rosly  Poisson structure on the moduli space
of flat $P^\uparrow_3$-bundles
one requires an element $r\in so(2,1)\otimes so(2,1)$
which satisfies  the classical Yang-Baxter equation and is such
that its symmetric part agrees with the non-degenerate invariant form
$\langle\,\,,\,\,\rangle$ used
in the definition of the Chern-Simons action (\ref{CSaction}).
One checks that
\bea
\label{rmatrix}
r= P_a \otimes J^a
\eea
satisfies the classical Yang-Baxter equation.
The symmetrised part
\bea
r^s = \frac{1}{2}( P_a\otimes J^a
+ J_a\otimes P^a)
\eea
 is  the invariant form on 
$iso(2,1)$ which we used in defining (\ref{CSaction}). 
As explained in \cite{schroers}, this implies that
 $r$ defines a  bi-algebra structure on  $iso(2,1)$
which  is co-boundary and quasi-triangular.


We do not enter further into a discussion  of the 
Poisson structure on the multi-particle phase space here,
which requires separate and detailed study. Instead we 
adopt the general philosophy of the combinatorial quantisation
programme of Chern-Simons theory, see \cite{AGSI}, \cite{AGSII}, 
\cite{AS} and \cite{schroers}. Roughly speaking, 
 the classical $r$-matrices play the role of 
structure constants in the Poisson structure of classical phase
space. To quantise, one looks for a quantum $R$-matrix which
has (\ref{rmatrix}) as a classical limit. This $R$-matrix then
plays the role of structure constants of the quantum algebra
of observables. In the next section we show that  a certain
deformation of the group algebra of the Poincar\'e group provides
the required $R$-matrix.

\section{The Lorentz double}

\subsection{Definition}
The quantum double of a group $H$ is a quasi-triangular Hopf
algebra constructed, via Drinfeld's  double construction, out of
the Hopf algebra of functions on $H$. As a vector space  the
quantum double  $D(H)$ is the tensor product of the  algebra of
functions $F(H)$ on $H$ and the group algebra $\CC(H)$. For finite
groups this definition makes rigorous sense, but when generalising
the construction to locally compact Lie groups $H$, one has to
make mathematical sense of the group algebra. Here, different
choices are made by different authors. The Lorentz double is the
quantum double $D(\LLor)$  constructed from the universal cover $\LLor$  of 
the orthochronous Lorentz group. In the definition we give below  we follow the
approach taken in \cite{KM} but generalise it slightly in the way
explained in \cite{schroers} for  the case $H=SU(2)$.

Before delving into technical details it is perhaps useful to give
a general description of the Lorentz double. As we shall explain,
it can be thought of as a deformation of  the group algebra of the
universal cover $\tilde P_3^\uparrow$   of the 
orthochronous Poincar\'e group. More precisely, there is a 
homomorphism from $D(\LLor)$ to the group algebra of 
$\tilde P^\uparrow_3$. However the co-algebra structure of $D(\LLor)$
differs from that of $\tilde P_3^\uparrow$. Physically this difference
corresponds to the non-commutative addition rule for momenta in
(2+1)-dimensional gravity.

The  easiest way to make sense of the group algebra of
$\tilde L_3^\uparrow$ is to consider generalised functions on
$\tilde L_3^\uparrow$ and to define the product via convolution:
 \bea
f_1*f_2(u)=\int_{\LLor} dv f_1(v)f_2(v^{-1}u).
\eea
As explained in
\cite{schroers} an  appropriate class  of generalised functions is
the set $M(\tilde L_3^\uparrow)$ of measures which are absolutely
continuous with respect to the Haar measure $dv$ on $\tilde L_3^\uparrow$
or pure point measures. Such measures can always be written 
in terms of an $L^1$-function and an $l^1$-sequence 
$\lambda_i$, $i\in \NN$
as
\bea
\label{meadecomp}
 dm = (f + \sum_{i\in
\NN}\lambda_i\delta_{h_i})dv.
\eea
 In practice this means that we can think of the elements of
$M(\tilde L_3^\uparrow)$ as $L^1$-functions, or Dirac $\delta$-functions
or a linear combination of both. It is advantageous to  include
Dirac $\delta$-functions  because the unit element is represented
by $\delta_e$.

The function algebra of $\tilde L_3^\uparrow$ is represented by a
different class of functions on $\tilde L_3^\uparrow$. This time the
multiplication is defined by pointwise multiplication of the
functions. For this to make sense, the functions should at least be
 continuous.  In practice we take  bounded, uniformly
continuous functions $C_B(\tilde L_3^\uparrow )$ on $\tilde L_3^\uparrow$.

Putting the ingredients together we define the Lorentz double
$D(\tilde L_3^\uparrow)$ as the set of  generalised functions $F$ on
$\tilde L_3^\uparrow\times \tilde L_3^\uparrow $ 
which are bounded and uniformly
continuous in the first argument   and measures of the form
(\ref{meadecomp}) in the second argument. We define a norm for
such  functions
 \bea
||F||_1=\int_{\LLor} dv \,\,\mbox{sup}_{h\in \LLor}|F(h,v)|
 \eea 
and equipped with that norm the Lorentz double becomes a Banach
algebra.

 We refer the reader to \cite{KM} and \cite{BM} for a
complete list of how the algebraic operations of a
quasi-triangular Hopf-$*$-algebra are implemented in $D(H)$.
For our purposes we only need the
formulae  for the multiplication, the co-multiplication, the ribbon
element  and the
universal $R$-element. The
multiplication of two elements $F_1$ and $F_2$ of $D(\tilde L_3^\uparrow)$
is
\bea 
\label{multi} \bigl(F_1\bullet F_2\bigr)(h,u)=\int_{\Lor}\,
dw\, F_1(h,w)F_2(w^{-1}hw,w^{-1}u).
\eea
The co-multiplication
$\Delta$ is defined via
\bea \label{comulti} \Delta
F(h_1,u_1,h_2,u_2) = F(h_1h_2,u_1) \delta_{u_1}(u_2). 
\eea
Finally  the central  ribbon element is 
\bea
\label{crib}
c(h,u)=\delta_h( u)
\eea
and the universal R-element is 
\bea
\label{bigrmatrix}
R(h_1,u_1,h_2,u_2)=\delta_{h_1}(u_2)\delta_e(u_1). 
\eea
Together they satisfy the spin-statistics relation
(\ref{spinstatrel}), as required for a ribbon Hopf algebra.

\subsection{The Lorentz double as a deformation of the $\tilde P^\uparrow_3$
group algebra}

These formulae may seem  complicated and far removed from the
Poincar\'e group $P_3^\uparrow$ which entered the classical
formulation of (2+1)-dimensional gravity. In order to exhibit the
intimate relation between $\tilde P_3^\uparrow$ and the Lorentz double
$D(\tilde L_3^\uparrow)$ we need to express the algebraic structure of
the  Poincar\'e group in a somewhat unusual fashion. Recall the
multiplication law (\ref{covermulti})
for two elements $(\ba_1,u_1 ),
(\ba_2,u_2)\in \tilde P_3^\uparrow$. 
The idea is to
consider the group algebra, written in terms of suitable functions
on the group, and then to perform a Fourier transform.
As a manifold
 $\tilde P_3^\uparrow \simeq \RR^3\times\tilde  L_3^\uparrow$, and the Haar
measure 
on $\tilde P_3^\uparrow$ is the product of the Lebesgue measure on
$\RR^3$ and the Haar measure on $\tilde L_3^\uparrow$. Concretely, if
$(\ba,u)\in \tilde P_3^\uparrow$ then we use the Haar measure
$d^3\ba\,du$. We  realise the group algebra of $\tilde P_3^\uparrow$
as the set $M(\tilde P_3^\uparrow)$ of bounded
 measures on $P_3^\uparrow$ either absolutely continuous with
respect to $d\ba\, du$ or pure point measures.   Again representing such
measures by generalised functions  $\hat f_1,\hat f_2$ (possibly
including delta-functions) the multiplication rule is the
convolution 
\bea (\hat f_1\hat \bullet\hat  f_2) (\ba,u)
=\frac{1}{2\pi^3} \int_{\RR^3\times \LLor} d^3b\,dw\,\hat
f_1(\bob,w)\hat f_2(\Lambda(w^{-1})(\ba-\bob),w^{-1}u). 
\eea 
Now we
perform  a Fourier transform on the first argument of $\hat f$,
thus obtaining a function $f$ on $(\RR^3)^*\times \tilde  L_3^\uparrow$:
\bea 
\label{halff}
 f(\bk,w)=\frac{1}{(2\pi)^3}\int_{\RR^3}d^3a\,
 \exp({-i\bk\cd\ba})\hat f(\ba,w).
\eea
Note that elements $\bk$ of the dual space  $(\RR^3)^*$ have the dimension of
inverse length. 

The Fourier transform of a bounded measure on $\RR^3$  is
a bounded,  uniformly continuous function on
 $(\RR^3)^*$. After applying the Fourier transform (\ref{halff})
 to functions in $M(\tilde P_3^\uparrow)$ we thus obtain
  generalised
functions  $f$ on $(\RR^3)^*\times \tilde L_3^\uparrow$ which are
bounded, uniformly continuous functions of the first argument and
measures of the form (\ref{meadecomp}) with respect to the second.
We denote this set by $\CC(\tilde P_3^\uparrow)$. The product of two
generalised functions $f_1,f_2$ in $\CC(\tilde P_3^\uparrow)$ is obtained
by applying the
 Fourier transform (\ref{halff}) to the convolution
product. The result is 
\bea 
\label{isomult} 
\bigl(f_1\bullet
f_2\bigr) (\bk,u)=\int_{\LLor}\,dw\, f_1(\bk,w)
f_2(\Lambda(w^{-1})\bk,w^{-1}u). 
\eea
The group-like co-multiplication for $P_3^\uparrow$ leads to  the
following  co-multiplication  for $f\in \CC(\tilde P_3^\uparrow)$   
\bea
\label{isocomult} (\Delta
f)(\bk_1,u_1,\bk_2,u_2)=f(\bk_1+\bk_2,u_1)\delta_{u_1}(u_2). 
\eea

It is now not difficult to write down promised
relationship between $D(\tilde  L_3^\uparrow)$ and $\tilde
P_3^\uparrow$. The idea is to interpret the first argument
of an element $F\in D(\tilde  L_3^\uparrow)$ as a group
valued momentum. To implement this we 
need   a family of exponential maps from the Lie algebra
$so(2,1)$ to $\tilde  L_3^\uparrow$. Let $J^\kappa_a=\kappa J_a$
for a real, positive parameter  $\kappa$, so that 
\bea
[J^\kappa_a,J^\kappa_b] = \kappa \epsilon_{ab}^{~~c} J^\kappa_c. 
\eea
Then define 
\bea 
\label{expfun} 
\widetilde{\exp}_\kappa: (\RR^3)^*\rightarrow
\tilde L_3^\uparrow,\quad 
\widetilde{\exp}_{\kappa}(\bk)=\widetilde{\exp}(k_aJ^\kappa_a),
\eea 
where $\widetilde{\exp}$ is the exponential map defined in
(\ref{expo}).
For the exponential map to make sense, its argument needs to be
dimensionless.
Since $\bk$ has the dimension of inverse length we require $\kappa$
to be of dimension length. The available constants $\hbar$
and $G$ can be combined in an essentially unique fashion to
give the dimension length. We set
\bea
\kappa =8\pi G\hbar
\eea
and shall see in the next section that 
\bea
\label{wavevector}
\bp=\kappa\bk
\eea
is physically interpreted as momentum measured in units of 
$8\pi G$.

 We  use the  map (\ref{expfun})  to define 
\bea \mbox{EXP}_\kappa^* :D(\tilde
L_3^\uparrow)\rightarrow \CC(\tilde P_3^\uparrow) 
\eea 
by
pull-back on the first argument: 
\bea\mbox{EXP}_\kappa^*(F)
(\bk,u)=F(\widetilde{\exp}_\kappa(-\bk),u).
\eea
This establishes the promised relationship between the Lorentz
double and the Poincar\'e group algebra. As explained
in the Euclidean setting in \cite{schroers}
EXP$_\kappa^*$ is an algebra homomorphism but not a homomorphism
of Hopf algebras. Rather one should think of 
$D(\tilde L_3^\uparrow)$ as a
 non-co-commutative deformation of $\CC(\tilde P_3^\uparrow)$.
The details of the argument given in \cite{schroers} can be 
adapted to the present situation by switching from the  Euclidean
metric to the Lorentzian metric $\eta$. The basic  point is that the 
pull-back of the $R$-matrix
\bea
\label{rkappa}
R_{\kappa}=\bigl(\hbox{EXP}^*_\kappa\otimes\hbox{EXP}^*_\kappa\bigr)( R )
\eea
is  a deformation of the classical $r$-matrix (\ref{rmatrix}) in
the sense that  
in the limit  $\kappa\rightarrow 0$
\bea
\label{result}
R_\kappa = 1\otimes 1 + i \kappa\,  r + \,{\mathcal O}(\kappa^2).
\eea
As explained at the end of the previous section, 
this result is an important
step in the combinatorial quantisation scheme.

\subsection{ The irreducible representations 
 of  the Lorentz double}

The irreps of 
 $D(\tilde L_3^\uparrow)$ are listed in
 appendix B, to which we refer the reader for our notational
conventions.
Here only we discuss the 
 elliptic representations in more detail. These
 are labelled by pairs $(\mu,s)$, where $\mu\in \RR^+$  labels
an elliptic  conjugacy class  in $\tilde L_3^\uparrow$  
and $s$ labels an irrep of the
centraliser $N_E$ of  the element $r(\mu )$ (\ref{rotations})
 in that conjugacy class.
The centraliser group is isomorphic to $\RR$,
so  all irreps are one dimensional and labelled by a (real-valued) 
spin $s \in \RR$. The elliptic representations  $\varpi_{\mu s}$ are
labelled by the mass parameter  $\mu\in \RR^+$ and the spin parameter
 $s \in\RR$. Their carrier space is the again the space $V_{\mu s}$
defined in  (\ref{Hilbert}),  which 
also carries the representations of $\tilde P_3^\uparrow$. 
The action of an element $F\in D(\tilde  L_3^\uparrow)$  on this space is
\bea 
\label{Drep}
(\varpi_{\mu s}(F)\phi) (x) = \int_{\tilde  L_3^\uparrow}\,dw\,
 F(x r(\mu) x^{-1},w)\phi(w^{-1}x).
\eea
We introduce the abbreviation 
\bea
\label{gmomentum}
g(\mu,x)=x r(\mu) x^{-1}
\eea
 and note that 
in terms of the  parametrisation (\ref{minkeulerr}) 
for  $L=\Lambda(x)$, it 
follows from  $r(\mu)=\widetilde{\exp}(-\mu J_0)$ 
that 
\bea
\label{gmomentumm}
g(\mu,x)=\widetilde{\exp}(-p_a J^a),
\eea
where $\bp$ is the classical momentum (\ref{firstmom}).
The first argument of $F$ should therefore be thought of as 
an exponentiated or group valued momentum. It then follows from 
(\ref{wavevector}) that the vector $\bk$ is the de Broglie wave vector 
associated to $\bp$.

It is now straightforward to calculate
 the action of the central ribbon  element $c$
(\ref{crib})
on any element $\phi \in V_{\mu s}$. Since  $c$ is central it acts
by a phase. A short calculation shows
\bea
\label{ceigenval}
\varpi_{\mu s}(c )\phi = e^{is\mu}\phi.
\eea

Just as in the case of the Poincar\'e representation,
one can use the space $L^2(H_\mu,dm)$ of square-integrable functions
on the mass hyperboloid as the carrier space of elliptic 
 irreps of the Lorentz
double. The action $\varrho(F)$ on an element $\Psi \in L^2(H_\mu,dm)$ 
is 
\bea
\label{Drepp}
\varrho_{\mu s}(F)\Psi(\bp)= 
\int_{\tilde  L_3^\uparrow}\,dw\,
 F(\widetilde{\exp}(-p_a J^a),w) \exp({is\psi(w,\bp)})\Psi(\Lambda(w^{-1})\bp).
\eea
where the   phase factor  $\exp({is\psi(w,\bp)})$  is defined via
(\ref{multiplierphase}).

Finally, consider tensor products of irreps of the Lorentz double,
studied in a more general context in \cite{KBM}. 
Here we only need the action of 
the universal $R$-element
(\ref{bigrmatrix}) in the tensor product representation
 $\varpi_{\mu_1 s_1}\otimes\varpi_{\mu_2 s_2}$   on  some state
$\Phi \in   V_{\mu_1 s_1}\otimes V_{\mu_2 s_2}$. It 
is
\bea
\label{Raction}
\bigl( \varpi_{\mu_1s_1}\otimes \varpi_{\mu_2 s_2} \,(R)\,\Phi \bigr)(x_1,x_2))
= \Phi(x_1,g^{-1}(\mu_1,x_1)x_2).
\eea
Thus the effect of $R$ is to rotate or boost  the group valued momentum
of the second particle by the group valued momentum of the first.
Similarly one computes the effect of the monodromy element
$Q=R_{21}R$:
\bea
\label{Qaction}
\bigl( \varpi_{\mu_1 s_1}\otimes \varpi_{\mu_2 s_2} \,(Q)\,\Phi \bigr)(x_1,x_2))
= e^{-i(s_1\mu_1+s_2\mu_2)}\Phi(g^{-1}x_1,g^{-1}x_2),
\eea
where $g$ is the total group valued momentum of the system:
\bea
g=g(\mu_1,x_1)g(\mu_2,x_2). 
\eea

\section{Quantum scattering of gravitating particles}

As outlined in the introduction, the strategy of this section
is to base our study of the quantum theory of  gravitating particles
in 2+1 dimensions on the analogy with the theory of 
ordinary anyon scattering developed in sect. 3. There the Hilbert
spaces of single and several anyons were constructed out of 
irreducible representations of the Poincar\'e group (external
or space-time symmetry) and  of an appropriate ribbon Hopf algebra
(internal symmetry). For gravitating particles  external and 
internal symmetry are both captured by the same 
object,  namely  the  Lorentz double. 
 It is not clear {\it a priori} that the bundle construction
of the multi-particle Hilbert spaces of sect. 3 have a well-defined
analogue in the  gravitational context. We shall now show that this is 
the case, and test our result against the scattering formulae of 
't Hooft \cite{Hooft}, Deser, Jackiw and Sousa-Gerbert \cite{DJ},
\cite{SGJ}. The following discussion is close in spirit to the paper
\cite{Carlipscat} on quantum scattering of gravitating particles.
We recover many of the results of that paper but our more
general framework allows us to include particles with spin.

\subsection{One- and two-particle Hilbert spaces}

As in  discussion  of ordinary anyons we begin by
looking at a single particle of mass $\mu$ and 
spin $s$. We restrict attention to $\mu\in(0,2\pi)$
in order to have a simple classical interpretation of 
the space-time surrounding the particle as described in
sect. 2.2. In the representation theory of $D(\LLor)$ there
is no need to restrict masses to lie in this range, a point
to which we return briefly at the end of this paper.
Since external and internal 
symmetry coincide in gravity, there is no need to 
introduce separate external and internal Hilbert spaces as 
in sect. 3. Instead we consider  the irrep $V_{\mu s}$ of 
the Lorentz double and 
impose the analogue of the spin  condition 
(\ref{spinquant}).   Formally this condition takes
the same form as for the relativistic anyons  in sect. 3,
namely the requirement that the two central elements $\Omega$ and  $c$  agree
when acting on any element in $V_{\mu s}$. Thus not all combinations 
of mass and spin are allowed for particles in (2+1)-dimensional gravity.
One-particle Hilbert spaces are of the  form
\bea
\label{onegrav}
{\cal H}_1= V_{\mu s } \,\,\,\mbox{provided}\,\,\,
\varpi_{\mu s}(\Omega)\phi =
 \varpi_{\mu s }(c)\phi \quad \forall \phi \in V_{\mu s }.
\eea 
 Using the expression (\ref{ceigenval})
for the eigenvalue of $c$ we find that  in gravity the quantisation
condition for the spin  depends on the  mass:
\bea
\label{gravspinquant}
e^{2\pi i s} = e^{i\mu s }\quad\mbox{or}\quad s=
{n \over 1-{\mu \over 2 \pi}},
\quad n\in \ZZ.
\eea
This formula provides the first  point of contact between
our algebraic approach to quantising (2+1)-dimensional gravity 
and some of the basic phenomena described in our introductory
sect. 2. We find that the spin of particle of mass $\mu$
satisfies the  same quantisation condition as the angular
momentum (\ref{lquant}) of a test particle moving in the conical 
space-time created by that  particle.

Next  consider two particles, with masses $\mu_1$ and $\mu_2$ 
and spin $s_1$ and $s_2$, both satisfying the spin quantisation
 condition (\ref{gravspinquant}).
 Thus there exist two
integers $n_1$ and $n_2$ such that 
\bea
\label{singlespin}
s_1=
{n_1 \over 1-{\mu_1 \over 2 \pi}},
\quad s_2=
{n_2 \over 1-{\mu_2 \over 2 \pi}}.
\eea 
In order to keep our  discussion of the interaction of these particles
as simple as possible we restrict attention to
$\mu_1,\mu_2\in(0,\pi)$ to start with.
In order to construct the two-particle Hilbert space we begin with
the space $L^2(\widetilde{\cal M}_2,\CC)$ of square-integrable
complex-valued functions on the cover  of the space of  two distinct
velocities (\ref{multishell}). We again write $\bq_1=\bp_1/\mu_1$
and $\bq_2=\bp_2/\mu_2$ for the individual velocities and $\tilde \bq$
for elements of the covering space $\widetilde{\cal M}_2$.
The  Lorentz double acts on elements
$\Psi \in L^2(\widetilde{\cal M}_2,\CC)$  via 
\bea
\label{tildeaction}
\varrho(F)\psi(\tilde{\bq}) =\int_{\tilde  L_3^\uparrow}\,dw\,
 F(g(\tilde \bq),w)
\exp(is_1\psi(w,\bp_1)  + is_2\psi(w,\bp_2))\Psi(w^{-1}\tilde{\bq}),
\eea
where 
\bea
\label{totalgroup}
g(\tilde\bq)=\widetilde{\exp}(-p_1^aJ_a)\widetilde{\exp}(-p_2^aJ_a)
\eea 
is the total group valued momentum  associated to the pair of
velocities covered by  $\tilde \bq$.

Not all elements of $L^2(\widetilde{\cal M}_2,\CC)$
are physically allowed. As in the case of anyons we demand
that  under the  (inverse) $2\pi$ rotation $\Omega^{-1}$ physical
states transform according to the monodromy operator $Q$. The formula
(\ref{Qaction}) defines 
the action of $Q$ on the tensor product of two one-particle irreps.
However,  there is a natural way to let  $Q$ act on 
$L^2(\widetilde{\cal M}_2,\CC)$ as well. Using the relation
(\ref{spinstatrel}) and the action (\ref{tildeaction}) we define
\bea
\label{losingit}
\varrho(Q)\Psi(\tilde{\bq} )=e^{-i(s_1\mu_1  + is_2\mu_2)}
\varrho(c)\Psi(\tilde{\bq} ).
\eea
Then  the requirement on physical two-particle states becomes
\bea
\varrho(Q)\Psi(\tilde{\bq}) =\Psi({\Omega^{-1}\tilde\bq}).
\eea
Since  both particles satisfy the spin quantisation
condition this is equivalent to
\bea
\label{thecond}
\varrho (c)\Psi = \varrho(\Omega)\Psi.
\eea
To solve this condition explicitly we change coordinates
from $\tilde q$ to a relative boost and an overall Lorentz
transformation, as in our discussion in sect. 3.

In order to keep the notation simple, we use the abbreviation
$g_1$ and $g_2$ for the group valued momenta (\ref{gmomentum}) 
of particle 1 and 2, i.e
\bea
\label{standard}
g_1=\widetilde{\exp}(-p_1^aJ_a)\qquad
g_2=\widetilde{\exp}(-p_2^aJ_a).
\eea
The total group valued momentum is the product $g_1g_2$.
Here  we encounter a   subtlety of momenta in (2+1)-dimensional
gravity. It is possible for the  
 total group valued momentum  not to be in an elliptic 
conjugacy class of $\LLor$. In order to understand this phenomenon
we use a Lorentz transformation
to move to the rest frame of particle 1. In that frame the group
valued momentum of particle 1 is the rotation element $g_1=r(\mu_1)\in 
\tilde L_3^\uparrow$. The second particle will in general be moving
in this frame, but using rotations we can assume that it is moving 
along the 1-axis. Then, with the notation
(\ref{boosts}),  the  group valued momentum of the second
particle in the rest frame of the first is
\bea
g_2=b(\xi)r(\mu_2)b^{-1}(\xi)=\widetilde{\exp}(-\mu_2\cosh\xi J_0
-\mu_2\sinh\xi J_1).
\eea
As first discussed by Gott in \cite{Gott}  the total
group valued momentum fails to be in an elliptic class  when  the 
second particle is  moving too fast relative to the first.
Quantitavely, the condition for an element $g\in \LLor$ to be in an elliptic
conjugacy class is $-1<\tr ( U(g))<1$. Applying this to the product
$g_1g_2$ and evaluating in the rest frame of particle 1 one finds
\bea
\label{Gottcond}
\cosh \xi\,\sin\frac {\mu_1}{2}\sin \frac{\mu_2}{2}<1+ 
\cos\frac {\mu_1}{2}\cos \frac{\mu_2}{2}.
\eea
We postpone the very interesting discussion of what happens
when this condition is violated. Assuming that it is satisfied 
we can find  a Lorentz transformation $w(\xi)$ such that 
\bea
\label{summ}
r(\mu_1) b(\xi)r(\mu_2)b^{-1}(\xi) = w(\xi)r(\mu(\xi))w^{-1}(\xi).
\eea
This condition defines the invariant mass $\mu(\xi)$  of the
two particle system. Under the assumption of the condition 
(\ref{Gottcond}) the range of the invariant  mass is restricted:
\bea
\label{massrest}
\mu_1+\mu_2 <\mu(\xi)<2\pi.
\eea
As in our discussion of special relativistic particles in sect. 3
the invariant mass  and  the  
Lorentz transformation $w$ both depend on  $\xi$, $\mu_1$ and $\mu_2$,
but we suppress the  dependence on  $\mu_1$ and $\mu_2$ in our
notation.  Note that  the function $\mu(\xi)$ is  different from the
(rescaled)  non-gravitational
invariant mass  $8\pi G M(\xi)$ defined in (\ref{totalmass}) but approaches
it in the limit of small relative speed
 $\xi \rightarrow 0$. Note also that, unlike $V$
in (\ref{sum1}) the element $w\in \LLor$ will not generally be 
a  boost along the $1$-axis. In other words, the momenta $\bp_1$
and $\bp_2$ in the centre of mass frame do not point in 
opposite directions in the sense of flat space physics. 
To see this explicitly,
consider  the condition of being
in  the centre of mass frame:
\bea
\label{comm}
g_1g_2=r(\mu)
\eea
for some given invariant mass $\mu$. Suppose that  particle
1 is moving along the 1-axis, i.e.
\bea
g_1=\widetilde\exp(-\mu_1\cosh\vartheta_1 J_0 -\mu_1\sinh\vartheta_1 J_1).
\eea
Then solving 
 the condition (\ref{comm}) for $g_2$ we find
\bea
g_2=\widetilde\exp(-\mu_2\cosh\vartheta_2 J_0-\mu_2\sinh\vartheta_2 
\cos \varphi J_1
-\mu_2\sinh\vartheta_2 \sin \varphi J_2)
\eea
with $\varphi=\pi+\mu/2$ and  the masses and rapidities related by
\bea
\cos\frac{\mu_2}{2}&=&\cos\frac{\mu_1}{2}\cos\frac{\mu}{2}
+\sin\frac{\mu_1}{2}\sin\frac{\mu}{2}\cosh\vartheta_1 \nonumber \\
\sin\frac{\mu_1}{2}\sinh\vartheta_1 &=&\sin\frac{\mu_2}{2}\sinh\vartheta_2.
\eea
In other words, in the centre of mass frame
the directions of motion
are related by a rotation by $\varphi=\pi-\mu(\xi)/2$ instead of the familiar
$\pi$. The momenta  thus behave as if they belonged to particles
moving in opposite directions on a cone of deficit angle $\mu(\xi)$.

 Despite these differences with the situation in sect. 3  we can again
  change variables from the momenta of  the particles
 to a relative boost parameter $\xi> 0$ and an overall Lorentz transformation.
We  define a new set of reference velocities
\bea
\label{rrelate}
\bq_1(\xi) =\Lambda(w^{-1})(\xi)(1,0,0)^t
\quad \mbox{and} \quad
\bq_2(\xi)=  \Lambda(w^{-1}(\xi)b(\xi))(1,0,0)^t
\eea
and the bijection 
\bea
\label{bk}
K: L^\uparrow_3 \times \RR^+ &\rightarrow& {\cal M}_2 \\
  (L,\xi)&\mapsto& (L\bq_1(\xi)  , L\bq_2(\xi)  )=(\bp_1,\bp_2).
\eea

With our assumption that  the group valued 
the total momentum is in an elliptic conjugacy class
 we can define a linear total momentum $\bP$ of the two particle
  system 
via
\bea
\widetilde{\exp}(-p_1^aJ_a)\widetilde{\exp}(-p_2^aJ_a)=
\widetilde{\exp}(-P^aJ_a).
\eea
In terms of  the
invariant mass $\mu(\xi)$ and the parameterisation (\ref{com}) 
of $L$ and we find  that the expression for  $\bP$ has the familiar form 
\bea
\label{comformulaa}
\bP=(\mu(\xi)\cosh\vartheta,\mu(\xi)\sinh\vartheta\cos\theta,
\mu(\xi)\sinh\vartheta\sin \theta).
\eea

For the application in the quantum theory  we again need 
to lift the map (\ref{bk}) to the covers $\LLor\times \RR^+$
and  $\widetilde{\cal M}_2$.
Imitating the steps leading up to (\ref{bilift}) we pick
a lift  $\tilde \bq(\xi) $ of the reference momentum
$(\bq_1(\xi),\bq_2(\xi))\in {\cal M}_2$.
Then, using the action of $\LLor$
on  $\widetilde{\cal M}_2$ we define the  bijection
\bea
\label{bjlift}
\tilde K: \LLor \times \RR^+ &\rightarrow&\widetilde{ {\cal M}_2} \\
  (u,\xi)&\mapsto& u\tilde \bq(\xi).
\eea
The crucial aspect of the map $K$ and its cover  is that it  preserves 
the total group valued momentum in the following sense. Suppose
$\tilde \bq= u {\tilde \bq}(\xi)$ covers the pair of velocity vectors
$(\bq_1,\bq_2)$ and that $\tilde K(u,\xi)=\tilde \bq$. The one checks
  that 
\bea
g(\tilde \bq)
=u r(\mu(\xi))u^{-1},
\eea
where $g(\tilde \bq)$ is the total momentum associated to $\tilde \bq$
via (\ref{totalgroup}).
At the level of representations this relation implies that the pull-back of
  elements of 
$L^2(\widetilde{M}_2,\CC)$  to $L^2(\tilde L_3^\uparrow\times
  \RR^+,\CC)$
via  $\tilde K^* $ intertwines between 
 the Lorentz double action 
\bea
\Pi(F)\Phi(x,\xi)=\int_{\tilde  L_3^\uparrow}\,dw\,
 F(x r(\mu(\xi)) x^{-1},w)\Phi(w^{-1}x,\xi)
\eea
on $L^2(\tilde L_3^\uparrow\times \RR^+,\CC)$ and 
the action (\ref{tildeaction}) on $L^2(\widetilde{M}_2,\CC)$. 

Proceeding further along the lines of the anyon discussion in sect. 3
we use this interwiner to express our condition (\ref{thecond})
on physical two-particle states in terms of elements
of  $L^2(\tilde L_3^\uparrow\times\RR^+,\CC)$. Because of 
the intertwining property the condition takes the same
form
\bea
\label{thecondd}
\Pi(c)\Phi=\Pi(\Omega)\Phi.
\eea
The Hilbert space of physical   two particle states is thus 
\bea
\label{thespace}
{\cal H}_2=\{\Phi\in L^2(\tilde L_3^\uparrow\times\RR^+,\CC)|
\Pi(c)\Phi=\Pi(\Omega)\Phi\}.
\eea

The advantage of  expressing the condition (\ref{thecondd}) on
the  space 
$L^2(\tilde L_3^\uparrow\times\RR^+,\CC)$ is that it can now
be solved explicitly.
First we decompose $L^2(\tilde L_3^\uparrow\times \RR^+,\CC)$
into eigenspaces of the right action of rotations $r(\psi)$:
 \bea
V_{s}=\{\Phi\in L^2(\tilde L_3^\uparrow\times \RR^+,\CC)|
\Phi( x r(\psi),\xi)\Phi=e^{-i\psi s} \Phi(x,\xi)\,\forall
\psi \in \RR\}.
\eea
{\it A priori} the eigenvalue $s$ can take arbitrary real values,
so we have the direct integral decomposition
\bea
L^2(\tilde L_3^\uparrow\times \RR^+,\CC)=\int_{s\in \RR} ds  \,V_s.
\eea
Then note that for $ \Phi\in V_{s}$ 
\bea
\Pi(c)\Phi=e^{i\mu(\xi)s}\Phi
\eea
and
\bea
\Pi(\Omega)\Phi=e^{2\pi i s}\Phi.
\eea
When  we impose the equality (\ref{thecondd})
we  select  those component spaces $V_s$ 
satisfying the spin selection rule
\bea
\label{spinsumruleg}
e^{2\pi i s}= e^{i\mu(\xi)s}.
\eea
While this rule looks similar to (\ref{spinsumrule}) there is 
an important difference. In gravity, the total spin of the 
two particle system is quantised in units which depend 
on the centre of mass energy of the two particles. Solving 
the condition we find that 
\bea
\label{explicitspin}
s=\frac n {1-\frac{\mu(\xi)}{2\pi}}, \quad n\in \ZZ.
\eea
The two-particle Hilbert space can thus be  decomposed into
irreps of the Lorenzt double as follows:
\bea
\label{Hilbertdecomp}
{\cal H}_2 = \int_{\mu_1+\mu_2}^{2\pi} d\mu 
\bigoplus_{s} V_{\mu s},
\eea
where, for given $\mu$, 
 we sum over all $s$ of the form (\ref{explicitspin}).
Each component space $V_{\mu s}$ thus satisfies the one-particle
spin quantisation condition. Recall that the range 
of the invariant mass in the direct integral is determined 
by our imposition of the Gott condition according to (\ref{massrest}).

Much of the physics of two gravitating particles is contained in
 the quantistation condition (\ref{explicitspin}). 
The orbital angular momentum $l$ in the centre of mass frame
of the two particles is related to $s$ via
\bea
\label{orbital}
l=s-s_2-s_2.
\eea
It follows that  for spinless particles  the orbital angular
momentum is quantised according to the rule
\bea
\label{lquantt}
l=\frac n {1-\frac{\mu(\xi)}{2\pi}}, \quad n\in \ZZ.
\eea
This rule generalises the formula (\ref{lquant}) and agrees 
with the quantisation condition found in \cite{Carlipscat}. The more 
general formula (\ref{orbital}) for particles with spin
implies
\bea
l=\frac n {1-\frac{\mu(\xi)}{2\pi}} -s_1-s_2 \quad n\in \ZZ,
\eea
which   can be 
written as 
\bea
\label{orbitalre}
l=\frac {\tilde{n}} 
{1-\frac{\mu(\xi)}{2\pi}} +\frac 1 {1-\frac{\mu(\xi)}{2\pi}}
\left(\frac{\mu(\xi)-\mu_1}{2\pi} \, s_1 + 
\frac{\mu(\xi)-\mu_2}{2\pi} \, s_2\right),
 \qquad \tilde{n} \in \ZZ.
\eea
with the simple shift $\tilde{n} =n -n_1- n_2$ by the integers characterising 
the individual spins according to (\ref{singlespin}).
This formula should be compared with various   conjectured
formulae for the quantistation of orbital angular momentum
of two gravitating particles with spin. In \cite{WC} the situation
of one spinning particle moving in the background of one heavy
particle is considered and in \cite{SGJ} the authors study the 
Dirac equation on a ``spinning cone''.
The quantisation rule implicit in  the (conjectured) general
scattering cross section found in \cite{SGJ} is
 \bea
\label{sgjcon}
l=\frac {\tilde{n}} {1-\frac{\mu}{2\pi}} +\frac 1 {1-\frac{\mu}{2\pi}}
\left(E_2 \, s_1 + E_1 \, s_2\right),
 \quad \tilde n \in \ZZ,
\eea
where $E_1$ and $E_2$ are the energies of particles 1 and 2.
We note that this formula and also the one conjuctured in
\cite{WC} contain the term (\ref{lquantt})
 with the spin-dependent correction
term containing  products of the form ``spin of one particle'' $\times$
``energy associated with the other particle''. Our formula is also 
of that form,  but the details are different. We stress that 
our result is obtained by a calculation 
 in a completely relativistic framework which treats both
particles on equal footing, whereas the 
the general formulae in \cite{SGJ} and \cite{WC}
are conjectured generalisations of results obtained
by studying the motion of a lighter particle in the background 
geometry due to a heavier particle.  Our result  agrees with 
the results of  \cite{SGJ} in the limiting case where 
one particle, say particle 1,
 is at rest and the second  particle is light compared to the first
 and moving slowly.
 In that case $E_1=\mu_1$ and  $\mu \approx \mu_1 +E_2$ 
so that $s_1(\mu-\mu_1)\approx s_1 E_2$. If $E_2-\mu_2$ small
compared to $\mu_2$ then we also have $s_2(\mu-\mu_2)\approx s_2 E_1$.
 The relative factor 
of  $2\pi$ is   due to  a different choice of units in \cite{SGJ}.

\subsection{Gravitational scattering revisited}

Exploiting futher our analogy with  non-gravitational anyons
in sect. 3 it is now not difficult to compute cross sections
for gravitational scattering of massive particles. One important
aspect of the calculations in sect. 3  which makes their adaptation
to gravity straightforward is that they were formulated entirely
in momentum space.  We shall see that gravitational scattering
in 2+1 dimensions can be studied and computed entirely
in momentum space.
As always  when comparing the outcome of a quantum mechanical
calculation with a (thought) experiment we need to use quantum
states which have a classical interpretation. Just as in sect. 3
these will be  (non-normalisable) momentum eigenstates.  Thus,
while we perform our calculation in the space (\ref{thespace})
it is essential for the interpretation of our result that we can
relate our computation to two particle momentum eigenstates via
(\ref{bjlift})
The two particle scattering states have the same form as 
in (\ref{twoscatt}):
\bea
|\bp_1,\bp_2\rangle=\frac{(8\pi^2)^2}{\mu_1\mu_2}
\frac{1} {\sinh \vartheta_1}\delta_{\vartheta_1}
\frac{1} {\sinh \vartheta_2}\delta_{\vartheta_2}
\delta_{\varphi_1}\delta_{\varphi_2}.
\eea
Following through the steps leading to (\ref{totalmom})
we write this state in terms of the total momentum $\bP$
(\ref{comformulaa}) and the angle $\varphi$ which gives the spatial
orientation  of the momenta in the centre of mass frame
(recall that the relative spatial angle
between the momenta in the centre of mass frame
is fixed to be $\pi-\mu(\xi)/2$):
\bea
\label{totalmomm}
|\bp_1,\bp_2\rangle=(8\pi^2)^2\mu(\xi)\delta_{P_0}
\delta_{P_1}\delta_{P_2}\delta_\varphi.
\eea
For gravitational scattering, the scattering states 
are completely characterised by the momenta. Fixing the relative
boost parameter $\xi$ and hence 
invariant mass $\mu(\xi)$ we introduce the abbreviation
\bea
\alpha(\xi)=1-\frac{\mu(\xi)}{2\pi}.
\eea
Scattering states have a definite total invariant mass but
contain all allowed spin states. They should be 
thought of as non-normalisable elements of the direct sum
\bea
\bigoplus_{s=\frac{n}{\alpha(\xi)},\, n\in\ZZ} V_{\mu(\xi)s}
\eea
appearing in the direct integral decomposition
(\ref{Hilbertdecomp}).
Scattering states can therefore be Fourier expanded as 
\bea
|\bp_1,\bp_2\rangle & =&4 (2\pi)^3M(\xi)\delta_{P_0}
\delta_{P_1}\delta_{P_2}\sum_{n \in \ZZ}e^{i\frac{n}{\alpha}\varphi}
|s\rangle.
\eea

The $S$-matrix  is again constructed according to the general
prescription (\ref{plekscat}). It is diagonal in each of the 
components $V_{\mu s}$ of the decomposition (\ref{Hilbertdecomp})
and its eigenvalues in the space $V_{\mu s}$ are 
\bea
S^{(l)} = 
\left\{ \begin{array}{cc}  e^{-\frac {i}{2}(\mu s -\mu_1 s_1-\mu_2
      s_2)} \,\,\, \mbox{if}  
&\quad [l]\geq 0 \\  e^{\frac {i}{2}(\mu s -\mu_1 s_1-\mu_2 s_2)}
\quad \mbox{if} & 
\quad [l]<0, \end{array} \right.
\eea
where again $l=s-s_1-s_2$.
Note that 
\bea
e^{i(\mu s-\mu_1s_1-\mu_2s_2)}=e^{2\pi i( s-s_1-s_2)}=
e^{2\pi il}
\eea
so that we have the simple formula
\bea
\label{gravs}
S^{(l)} =\left\{ \begin{array}{ll}&e^{-i\pi l} \quad \mbox{if}
\quad [l] \geq 0  \\
 &e^{i\pi l}\,\,\,\,\quad  \mbox{if}\quad
[l] < 0,  \\
\end{array} \right.
\eea
with all the physics in the quantisation condition for $s$ (and hence $l$).

Following the manupilations in sect. 3 we 
define the  reaction matrix $T$ and its  reduced matrix
element as in (\ref{sdecomp}) and (\ref{tred})
and find that it can be written as  
\bea
\langle\bp_1^f,\bp_2^f|{\cal T}|\bp_1^i,\bp_2^i\rangle
=4 \mu(\xi) t(\xi, \varphi^i -\varphi^f),
\eea
where the function $t(\xi, \varphi)$ encodes the dependence of the 
scattering on the scattering angle $\varphi=\varphi^i-\varphi^f$.
Note that, unlike its non-gravitational counterpart (\ref{tofphi})
it also depends on the rapidity $\xi$.
Before we discuss it in detail we show how it enters the expression for  
the differential cross section. Mimicking the derivation of phase 
space factors for anyon scattering in the calculation leading up 
to (\ref{hey}) we arrive at the following expression for the
differential cross section  in the centre of mass frame:
\bea
\label{heyh}
\frac{d\sigma}{d\varphi}(\varphi,\xi)
=\frac {\tilde M(\xi)\hbar }{2\pi\MM_1\MM_2\sinh\xi }|t(\xi, \varphi)|^2 ,
\eea
where we have re-instated $\hbar$ and written down the masses in
physical units i.e. $\mu_1=8\pi G \MM_1$, $\mu_2=8\pi G \MM_2$ and
 $\mu(\xi)=8\pi G \tilde M(\xi)$. Reverting to the masses expressed 
in units of $(8\pi G)^{-1}$ and recalling that the Planck length is 
$\ell_P=\hbar G$ we have the expression
\bea
\label{heyho}
\frac{d\sigma}{d\varphi}(\varphi,\xi)
=4 \frac{\mu(\xi) \ell_P }{\mu_1\mu_2\sinh\xi }|t(\xi, \varphi)|^2 .
\eea

Let us first compute the function $t(\xi,\varphi)$ for the case of spinless
particles. In that case $s=l$ and the analogue of the formula 
(\ref{univscat}) is  
\bea
i\,t(\xi, \varphi) =  \sum_l (S^{(l)} -
1)\,e^{il\varphi} 
=\sum_{n\in \ZZ} \left(e^{-i\frac{ \pi |n|}{\alpha(\xi)}}
-1\right) e^{i\frac{ n \varphi}{\alpha(\xi)}}.
\eea 
This leads to a scattering cross section (\ref{heyho})
 in agreement with the result of Deser and Jackiw
\cite{DJ} in the limit where one particle is very heavy
and the other very light (so that the invariant mass is
approximately equal to the mass of the heavy particle). 
The divergence in $t$ can be regularised in the way described 
in \cite{DJ} leaving the finite contribution
\bea
i \,\tilde{t}(\xi, \varphi) = 
\frac 1 2 \left(\cot\frac{\varphi-\pi}{2\alpha(\xi)} -
\cot\frac{\varphi+\pi}{2\alpha(\xi)}\right).
\eea
In our formalism it is easy to deal with spinning particles.
The amplitude function $t$ for particles of spin $s_1$ and 
$s_2$ is again given by the general expression (\ref{univscat}).
We find:
\bea
i\,t(\xi,\varphi) &=&
\sum_{n\in \ZZ} \left( e^{-i\pi|l|}-1\right)e^{is\varphi}\nonumber \\
&=&\sum_{n\in \ZZ} \left( e^{-i\pi|\frac{n}{\alpha}-s_1-s_2|}-1 \right)
e^{i\frac{n}{\alpha}\varphi}\nonumber \\
&=&\sum_{n\geq
  -[-\alpha(s_1+s_2)]}\left(e^{i\pi(s_1+s_2)}
e^{i\frac{n(\varphi-\pi)}{\alpha}} 
-e^{i\frac{n\varphi}{\alpha}}\right ) +  \nonumber \\
&&\sum_{n <
  -[-\alpha(s_1+s_2)]}\left(e^{-i\pi(s_1+s_2)}
e^{i\frac{n(\varphi+\pi)}{\alpha}}
 -e^{i\frac{n\varphi}{\alpha}}\right), 
\eea
where we have supressed the explicit dependence of $\mu$ and 
$\alpha$ on $\xi$ for notational simplicity.
The infinite sums an again be treated  as described in \cite{SGJ}.
The finite part is 
\bea
\label{scatresult}
i \tilde t(\xi,\varphi)& =&
\frac 1 2 e^{-i[\alpha(s_1+s_2)]\frac {\varphi }{\alpha}}
\left[ e^{-i\frac{\pi}{\alpha}\{-\alpha(s_1+s_2)\}}
\left(\cot\frac{\varphi-\pi}{2\alpha}-i\right)\right. \nonumber \\
&& -e^{i\frac{\pi}{\alpha}\{-\alpha(s_1+s_2)\}}\left.
\left(\cot\frac{\varphi+\pi}{2\alpha}-i\right) \right],
\eea
where  $\{x\}=x-[x]$ denotes the fractional part of $x$.
It is very interesting to compare this result with the conjectured 
universal scattering amplitudde given in the summary of \cite{SGJ}.
There is a trivial difference in kinematical factors which arises because
we work in the center of mass frame and our definition of $t$ is such
that it is stripped of all explicit momentum dependence. The formula
given in \cite{SGJ} is structurally similar to ours  and we can 
recover it in an appropriate limit  if we can relate our $\alpha (s_1+s_2)$
to  their $E_1s_2+E_2s_1$, where $E_1$ and $E_2$ are again the 
energies of particles 1 and 2.  The key to this relation lies 
again in the spin quantisation condition of the individual particles
(\ref{singlespin}). Since $(1-\frac{\mu_1}{2\pi})s_1$ and 
 $(1-\frac{\mu_2}{2\pi})s_2$ are both integers we have 
\bea
\{-\alpha(s_1+s_2)\}=
\left\{\frac{(\mu-\mu_1)}{2\pi} s_1+\frac{(\mu-\mu_2)}{2\pi} s_2\right\}.
\eea
Following the argument given in the discussion of the 
quantisation condition  (\ref{sgjcon}) this expression
 reduces  to $\{E_2s_1+E_1s_2\}$ 
in the limit where particle 1 is very heavy   and at rest and particle
2 is   a slow and  light test particle.

\section{Discussion and outlook}
For gravitating particles 
in 2+1 dimensions,  space-time symmetry
and internal symmetry are unified in  the Lorentz double.
Much of this paper has been devoted to justifying this statement
and to exploring its consequences. To end, we highlight and discuss
some of our key results.

The Lorentz double allows us to  give a precise a
description of the Hilbert space  for  the gravitational
system consisting of one or several 
gravitating particles in 2+1 dimensions. It does so in 
a way which generalises and makes precise various notions
which have appeared  in the literature on (2+1)-dimensional gravity.
For example, our definition of the  one-particle Hilbert space
(\ref{onegrav}) captures the idea, expressed e.g. in \cite{Carlipscat},
 that one should consider irreps
of the {\it universal cover} of the Poincar\'e group and pick out the 
physically allowed spin values  by a suitable mass-dependent condition.
 The ribbon element of the Lorentz double gives a precise algebraic
implementation of that idea.
For several particles, the role of the braid group in defining 
the Hilbert space has been anticipated in many publications, e.g.
 \cite{Carlipscat},
\cite{SG}, \cite{CCV1}, \cite{CCV2}, \cite{KMVW}, \cite{VW} and
\cite{KO}. Our formulation (\ref{thespace}) is in the spirit
of these papers. However, drawing on the analogy with ordinary anyons
and  using the representation of the braid group
furnished by the Lorentz double we are able to give a completely
general and invariant  definition of  the  Hilbert space of 
 gravitating particles with spins. 

Finally 
we have shown that the interaction of gravitating particles 
in 2+1 dimensions
can be computed by viewing them as gravitational anyons.
The interaction of ordinary anyons is dictacted by the universal
$R$-matrix of a suitable ribbon Hopf algebra and here we have
seen that, by drawing a careful analogy,
 the interactions of gravitating particles can 
be computed from the $R$-matrix of the Lorentz double. 
The results agree with older semi-classical calculations in certain
limits  and 
generalise them in a simple and very natural fashion. 
In particular we claim that the combination of (\ref{scatresult})
with (\ref{heyho}) gives a completely general formula for the 
gravitational 
scattering cross section of two spinning particles in the 
centre of mass frame, provided their relative speed satisfies
the condition (\ref{Gottcond}).

It is 
clear that the formalism developed here is powerful enough
to handle more general situations than we have discussed.
It would be interesting, for example, to study the scattering
when the relative speed violates the condition (\ref{Gottcond})
and to investigate the interpretations of elliptic representations
of the Lorentz double when the total mass exceeds $2\pi$.
Also, our computations could be extended to deal with several
particles, possibly moving in universes with non-trivial topology.

It remains an interesting challenge to give a detailed derivation
of  our results  
 in the framework
of combinatorial quantisation of Chern-Simons theory.
 Recent work on the  combinatorial quantisation
 of Chern-Simons theory with gauge group $SL(2,\CC)$
(corresponding to Lorentzian gravity with a cosmological positive
constant)
in \cite{BuRo} and  \cite{BNR} shows how these question can be addressed
but also illustrates the level of technical difficulty involved.

\vspace{1cm}

\vbox{
\noindent{\bf Acknowledgements}

\noindent
This paper has had a gestation period of several years, 
following a tentative treatment in \cite{Muller}. During that time 
 we  discussed the issues addressed here with many people.
We thank in particular Herman Verlinde for discussions during the 
early stages of the project and Tom Koornwinder for frequent 
mathematical tutorials on non-compact quantum groups. 
BJS acknowledges financial support through an Advanced Research
Fellowship of the Engineering and Physical Sciences Research Council.
}

\appendix

\section{The universal cover of the Lorentz group in 2+1 dimensions}

Our description of the universal cover of $L_3^\uparrow$
follows essentially that given in \cite{MS}, which in turn
follows \cite{Bargman}.  As an intermediate step
it is best to consider the double cover $SU(1,1)$. This group is a
subgroup of $SL(2,\CC)$, consisting of all $U\in SL(2,\CC)$ satisfying
\bea
U^{-1}=\pmatrix{1 &0 \cr 0  &-1}U^\dagger\pmatrix{1 &0 \cr 0  &-1}.
\eea
Elements can be written as
\bea
\label{albema}
U= \pmatrix{\alpha & \beta \cr \bar \beta &\bar  \alpha}
\eea
with the condition
\bea
\label{condi}
|\alpha|^2-|\beta|^2 =1.
\eea
For later use we also introduce  generators for the Lie algebra of $SU(1,1)$:
\bea
T_0=\pmatrix{-\frac i 2 & 0 \cr 0&\frac i 2},\quad
T_1=\pmatrix{ 0 & -\frac i 2 \cr \frac i 2 &0},\quad
T_2=\pmatrix{ 0 & \frac 1 2 \cr \frac 1 2 &0}.
\eea
They satisfy
\bea
T^\dagger_a=- \pmatrix{1 &0 \cr 0  &-1}T_a\pmatrix{1 &0 \cr 0  &-1}
\eea
and the algebraic relation
\bea
T_aT_b=-\frac 1 4 \eta_{ab}+ \frac  1 2 \epsilon_{abc} T^c,
\eea
where indices are raised with the Minkowski metric $\eta_{ab}=
\mbox{diag}(1,-1,-1)$  and $\epsilon_{012}=1$. A useful corollary
is the following formula for the exponential
\bea
\exp(-p^aT_a)=\cos\frac{\mu}{2}-\frac{2}{\mu}\sin\frac{\mu}{2}\,\,p^aT_a,
\eea
where $\bp$ is a time-like vector with $\bp^2=\mu^2$.
We also deduce  the commutation relations
\bea
[T_a,T_b]=\epsilon_{abc}T^c
\eea
and  the normalisation 
\bea
\tr \left(T_a T_b\right)=-\frac 1 2 \eta_{ab}.
\eea

The adjoint action of $SU(1,1)$ on its Lie algebra gives an
explicit homomorphism $\Lambda: SU(1,1) \rightarrow L_3^\uparrow$.
More precisely,  we define the $3\times 3$ matrix $\Lambda(U)$ via
\bea
UT_aU^{-1}=\sum_{b=0}^2\Lambda(U)_{ba} T_b
\eea
and find, in terms of the parametrisation (\ref{albema}),
\bea
\label{explor}
\Lambda(U)=\pmatrix{
|\alpha|^2 + |\beta|^2 &- 2\Rea(\alpha \bar \beta) & 2 \Ima(\alpha \bar \beta)
 \cr
-2 \Rea(\alpha \beta) & \Rea(\alpha^2 + \beta^2) &-\Ima(\alpha^2-\beta^2) \cr
-2 \Ima(\alpha\beta) & \Ima(\alpha^2 + \beta^2)& \Rea(\alpha^2- \beta^2)}.
\eea
We define  generators $J_a$ of the Lie algebra
$L_3^\uparrow$,  denoted  $so(2,1)$, via
\bea
J_a=\left. \frac d {d\epsilon}\right|_{\epsilon=0} \Lambda(\exp(\epsilon T_a)).
\eea
These are used in the main text, where  $so(2,1)$ is physically 
interpreted as momentum space.

There  are other of ways of parametrising $SU(1,1)$ matrices which 
are used in the main text. One  set of parameters  is analogous to the 
Euler angles used for $SU(2)$ matrices. Since elements of the 
form $\exp(-\varphi T_0)$  correspond  to (anti-clockwise) rotation
in the $p_1p_2$-plane and elements of the form $\exp(-\vartheta T_2)$
correpond to finite boosts along the $1$-axis, we can parametrise
a general element $U\in SU(1,1)$ in terms of two angles $\varphi \in[0,2\pi),
\psi \in [0,4 \pi)$ and a boost parameter $\vartheta\in [0,\infty)$ via 
\bea
\label{minkeuler}
U(\varphi,\vartheta,\psi)= e^{-\varphi T_0}e^{-\vartheta T_2}e^{-\psi T_0}.
\eea
Clearly there is a corresponding parametrisation of $\Lor$ matrices
\bea
\label{minkeulerr}
L(\varphi,\vartheta,\psi)= 
e^{-\varphi J_0}e^{-\vartheta J_2}e^{-\psi J_0}
\eea
in terms of angles $\varphi \in[0,2\pi),
\psi \in [0,2\pi)$ and a boost parameter $\vartheta\in [0,\infty)$. 

Another parametrisation  is useful for understanding
the universal cover of $SU(1,1)$.
Let $D=\{\gamma\in \CC| |\gamma| < 1\}$ be the open unit disk. Then
elements of $SU(1,1)$
can be parametrised in terms of $\gamma \in D$
and  $\omega \in [0,4\pi)$ via
\bea
\label{uparam}
U(\gamma,\omega)=(1-\gamma\bar \gamma)^{-{1\over 2}}
\pmatrix{ e^{{i\over 2}\omega} & \gamma e^{{i\over 2} \omega}\cr
\bar \gamma e^{-{i\over 2} \omega} & e^{-{i\over 2} \omega} }.
\eea
The universal cover $\tilde L_3^\uparrow$  can be identified with the
set
\bea
\left\{(\gamma,\omega)| \gamma \in D, \,\, \omega
\in  \RR \right\}.
\eea
The group multiplication is 
$(\gamma_1,\omega_1)(\gamma_2,\omega_2)=(\gamma_3,\omega_3)$
where
\bea
\gamma_3&=& {\gamma_2 + 
\gamma_1e^{-i\omega_2}\over 1+\gamma_1\bar \gamma_2e^{-i\omega_2}}
\nonumber \\
\omega_3 &=& \omega_1 + \omega_2 +{1 \over i}\ln
\left({1+\gamma_1\bar \gamma_2e^{-i\omega_2}  
\over 1+\bar \gamma_1\gamma_2e^{i\omega_2}}
\right)
\eea
with the logarithm  defined in terms of the power series.
We use small latin letters $u,v,w,x, ...$ to denote elements
of $\tilde L_3^\uparrow$.
We also introduce special names for elements
of two subgroups of $\tilde L_3^\uparrow$, 
both isomorphic to $\RR$. We write
\bea
\label{rotations}
r(\mu)=(0,\mu), \quad \mu\in \RR,
\eea
for elements which get mapped to (anti-clockwise, i.e. mathematically
positive)  spatial rotations $\exp(-\mu J_0)$
under the map
(\ref{hom2}) and 
\bea
\label{boosts}
b(\xi)=(-\tanh \frac{\xi}{2}, 0), \quad \xi \in \RR,
\eea
for elements which get mapped to boosts $\exp(-\xi J_2)$ along the 
1-axis under 
(\ref{hom2}).

In the main text we also require the exponential map from the Lie
algebra $so(2,1)$ to $\tilde L_3^\uparrow$. Since $\tilde L_3^\uparrow$
is not a matrix group, this exponential map cannot be written
in terms of a power series of matrices, but has to be defined
in  differential geometric terms, as for example in \cite{Varadarajan}.
We denote the exponential map by 
\bea
\label{expo}
\widetilde\exp: so(2,1)\rightarrow \tilde L_3^\uparrow
\eea
and note that 
\bea
\widetilde{\exp}(-\mu J_0)=r(\mu),\quad \mbox{and} \quad
 \widetilde{\exp}(-\xi J_2)=b(\xi).
\eea
As we shall discuss in detail in appendix B, 
this map is neither  injective nore surjective.

One checks that the map
\bea
\label{hom1}
U: \tilde L_3^\uparrow\rightarrow SU(1,1), \quad
(\gamma,\omega) \mapsto  U(\gamma,\omega),
\eea
where $ U(\gamma,\omega)$ is defined as in (\ref{uparam}),
defines a group homomorphism with kernel $\{(0,4\pi n)|n\in \ZZ \}\simeq \ZZ$.
Finally defining $\Lambda(\gamma,\omega):= \Lambda(U(\gamma,\omega))$
we have the group homomorphism
\bea
\label{hom2}
\Lambda: \tilde L_3^\uparrow\rightarrow L_3^\uparrow, \quad
(\gamma,\omega) \mapsto  \Lambda(\gamma,\omega)
\eea
with kernel $\{(0,2\pi n)|n\in \ZZ\}\simeq \ZZ$. This kernel is 
in fact the centre of $\tilde L_3^\uparrow$. Its generator
 plays an important role in our discussion. It is
 the rotation by $2 \pi$, for which we introduce a special name
\bea
\label{rotgen}
\Omega =r(2\pi).
\eea


\section{Irreducible representations of $D(\tilde L_3^\uparrow)$}

As explained in the main text,
irreps of $D(\tilde L_3^\uparrow)$
 are labelled by conjugacy classes in $\tilde L_3^\uparrow$
together with centraliser representations. Here we list the 
conjugacy classes and the centraliser groups of specified elements
in the conjugacy class. We give a physical interpretation 
of the conjugacy classes and  point out whether they  lie in the 
image of the exponential map (\ref{expo}).
The classification of the conjugacy classes in $SL(2,\RR)$
can be found in \cite{KM}. Since $SL(2,\RR)$ is conjugate
to $SU(1,1)$ in $SL(2,\CC)$, one can translate the list
given in \cite{KM} into a list of $SU(1,1)$ conjugacy classes,
from which the conjugacy classes in $\tilde L_3^\uparrow$
can be deduced.

\newpage

\parindent 0pt

{\bf  1. Elliptic Representations}

These are labelled by elliptic conjugacy classes,
consisting  of Lorentz transformations which are conjugate
to a  pure rotation (\ref{rotations}).
There is family  of  elliptic conjugacy classes labelled by the
integers:
\bea
E^n_\mu=\{x \Omega^n r(\mu) x^{-1}|x \in\tilde L_3^\uparrow\}, \quad  
0  <\mu <2\pi,\,\,\,  n\in \ZZ,
\eea
where $\Omega$ is the $2\pi$ rotation introduced in (\ref{rotgen}).
The elliptic conjugacy classes can all be reached by exponentiating
 time-like momenta $p^0J_0+p^1J_1+p^2J_2$ 
whose  invariant mass $|\bp|$ is not an integer multiple of $2 \pi$.
More precisely, defining
\bea
{\cal E}^n_\mu=\{p^0J_0+p^1J_1+p^2J_2\in so(2,1)|\bp^2=(2\pi n+\mu)^2,\,\,
\mbox{sign}(p_0)=-\mbox{sign}(n)\},
\eea
where we take sign$(0)$ to be $+$, 
  the exponential map 
(\ref{expo}) maps ${\cal E}^n_\mu$ bijectively onto $E^n_\mu$.

The centraliser of the $\tilde L_3^\uparrow$ action on
 $r(\mu)$ is the same for all $\mu\neq 2\pi n$. It is
the subgroup
\bea
N_E=\{r(\omega) \in \tilde L_3^\uparrow|\omega \in \RR\}.
\eea
This is the universal cover of the group of spatial rotations and
 is isomorphic to $\RR$.

\vspace{0.3cm}

{\bf 2.  Hyperbolic  representations}

These are labelled by hyperbolic conjugacy classes
consisting  of elements conjugate to a pure boost of the form 
(\ref{boosts}).
 There is a  family of hyperbolic  conjugacy classes, again indexed
by the integers
\bea
H^n_\xi=\{x \Omega^n b(\xi) x^{-1} |x\in\tilde L_3^\uparrow\}, 
\quad \xi \in \RR^+.
\eea
The hyperbolic classes $H^0_\xi$, $\xi\in \RR^+$, can be obtained
by exponentiating $so(2,1)$ elements corresponding to 
 space-like momenta $p^0J_0+p^1J_1+p^2J_2$,
$\bp^2=-\xi^2$. The hyperbolic classes $H^n_\xi$ for $n\neq 0$ are 
not in the image of the exponential map (\ref{expo}), showing in
particular that the exponential map is not surjective.

The centraliser of the $L_3^\uparrow$ action on
 $ b(\xi)$ is the same for all $\xi\in \RR^{>0}$. It is
the subgroup
\bea
N_H=\{(\tanh \xi,2\pi n) \in \tilde L_3^\uparrow|\xi\in \RR, n\in \ZZ\}.
\eea
Physically this is the group of boosts along the 1-axis and of
spatial rotations by an integer multiple of $2\pi$. It is isomorphic to
$\RR\times \ZZ$.

\vspace{0.3cm}

{\bf  3. Parabolic representations}

Consider  the  two  elements
\bea
\label{nplus}
v_+=\left(\frac{1+i} 2, -\frac \pi 2\right)
\eea
and
\bea
\label{nminus}
v_-=\left(\frac{-1+i} 2, \frac \pi 2\right).
\eea
Then define the conjugacy classes
\bea
P_+^n=\{x\Omega^nv_+x^{-1}| x\in \tilde L_3^\uparrow\},
\eea
and
\bea
P_-^n=\{x\Omega^n v_-x^{-1}| x\in \tilde L_3^\uparrow\}.
\eea
The conjugacy class $P_+^0$  can be obtained by 
exponentiating $so(2,1)$ elements corresponding to 
light-like momenta $p^0J_0+p^1J_1+p^2J_2$,
$\bp^2=0, \bp_0 >0$ and 
the conjugacy class $P_-^0$  can be obtained by 
exponentiating light-like momenta $p^0J_0+p^1J_1+p^2J_2$,
$\bp^2=0, \bp_0 <0$. Physically, they thus correspond to the 
exponentiated forward and backward light cones.
  The hyperbolic classes $P^n_\xi$ for $n\neq 0$ are 
not in the image of the exponential map (\ref{expo}), showing again
that the exponential map is not surjective

The centraliser of the $\tilde L_3^\uparrow$
action on
both  $v_+$  and  $ v_-$ is
the subgroup
\bea
N_P=
\left\{\left({\lambda \over 1+i\lambda},\arg(1-i\lambda) +2\pi n\right)
 \in \tilde L_3^\uparrow|\lambda \in \RR, n\in \ZZ\right\}.
\eea
Physically this group consists of  combined boosts and rotations
 and of rotations by integer
multiples  of $2\pi$. It, too,  is isomorphic to
$\RR\times \ZZ$.

\vspace{0.3cm}

{\bf 4. Vacuum representations}

Finally there are  classes, again indexed by the integers,
 consisting of one element each:
\bea
O^n=\{\Omega^n\}. 
\eea
Each of these one element classes is  the  image of the mass
 hyperboloids
\bea
\label{quantmass}
{\cal H}_n=\{p^0 J_0+p^1J_1+p^2J_2 \in so(2,1)|\bp^2= 2\pi |n|,\,
\mbox{sign}(p_0)=-\mbox{sign}(n)\} ,
\eea
 under the exponential map, showing 
that the exponential map  is not injective.

The centraliser of $O^n$  is the entire group $\tilde L_3^\uparrow$.


\newpage

\begin{thebibliography}{99}

\itemsep=\smallskipamount
\bibitem{Carlipbook} S.~Carlip, {\em Quantum gravity in 2+1 dimensions,}
Cambridge University Press, Cambridge, { 1998}.
\bibitem{Carlipscat} S.~Carlip, {\em Exact quantum scattering in 
(2+1)-dimensional gravity}, Nucl.~Phys. {\bf B324} (1989) 106--122.
\bibitem{BM} F.~A.~Bais and N.~M.~Muller, {\em  Topological field
 theory and the quantum double of} $SU(2)$,
Nucl.~Phys.  {\bf  B530} (1998) 349--400.
\bibitem{schroers}
B.~J.~Schroers, {\em Combinatorial quantisation of  Euclidean gravity
in three  dimensions}, pp 307-328 in 
 {\em Quantization of singular symplectic quotients}, eds
N.~P.~Landsman, M.~Pflaum and M.~Schlichenmaier,  Progress
in Mathematics {\bf 198}, Birkh\"auser 2001; {\tt math.qa/0006228}.
\bibitem{Fredenhagen} K.~Fredenhagen, {\em Sum rules for spin in 
(2+1)-dimensional quantum field theory}, in 
{\em Quantum groups}, Proceedings Clausthal 1989, eds H.~D.~Doebner,
J.~Hennig,  Lecture Notes in 
Physics {\bf 370}, Springer 1990.
\bibitem{FM}
J.~Fr\"ohlich and P.~A.~ Marchetti, {\em Spin-statistics theorem and 
scattering in planar quantum field theories with braid statistics},
Nucl.~Phys. {\bf B356} (1991) 533--573.
\bibitem{MS}
J.~Mund and R.~Schrader, {\em Hilbert spaces for nonrelativistic and
relativistic ``free'' plektons (particles with braid group statistics)},
 In {\em Advances in dynamical systems and quantum physics},
Proceedings,  Capri Quantum Physics 1993:0235-259;
{\tt  hep-th/9310054 }.
\bibitem{FR} V.~V.~Fock and A.~A.~Rosly, {\em Poisson structures on
moduli of flat connections on Riemann surfaces and $r$-matrices,}
 ITEP preprint {\bf  72-92}  (1992); see also {\tt math.QA/9802054}.
\bibitem{AGSI}
A.~Y.~Alekseev, H.~Grosse and V.~Schomerus, {\em Combinatorial
quantization of the Hamiltonian Chern-Simons Theory,}
Commun.~Math.~Phys. {\bf 172} (1995) 317--358.
\bibitem{AGSII}
A.~Yu.~Alekseev, H.~Grosse and V.~Schomerus, {\em Combinatorial
quantization of the Hamiltonian Chern-Simons Theory II,}
Commun.~Math.~Phys. {\bf 174} (1995) 561--604.
\bibitem{AS} A.~Yu.~Alekseev and V.~Schomerus, {\em Representation
theory of Chern-Simons observables,} Duke Math.~Journal {\bf 85}
(1996) 447--510.
\bibitem{Hagen}
C.~R.~Hagen, {\em The Aharonov-Bohm scattering amplitude}, Phys.~Rev
{\bf D41} (1990) 2015--2017.
\bibitem{Staru}
A.~Staruskiewicz, {\em Gravitation theory in three-dimensional space},
Acta Phys. Plon. {\bf 24} (1963) 735--740.
\bibitem{Hooft} G. 't Hooft, {\em Non-perturbative 2 particle scattering
amplitude in 2+1 dimensional quantum gravity}, Commun. Math. Phys. {\bf 117}
(1988) 685--700.
\bibitem{DJ} S. Deser and R. Jackiw, {\em Classical and quantum scattering
on a cone,} Commun. Math. Phys. {\bf 118} (1988) 495--509.
\bibitem{BR}
A.~O.~Barut and R.~Raczka, {\em Theory of group representations and 
applications}, World Scientific, Singapore 1986.
\bibitem{Grigore}
D.~R.~Grigore, {\em The projective unitary irreducible representations
of the Poincar\'e group in 1+2 dimensions}, J.~Math.~Phys. {\bf 37} (1996)
460--473.
\bibitem{CP} V.~Chari and A.~Pressley {\em Quantum Groups, }
Cambridge University Press,  Cambridge, {1994}.
\bibitem{FGR} K.~Fredenhagen, M.~R.~Gaberdiel and S.~M.~R\"uger,
{\em Scattering states of plektons (particles with braid statistics) in
2+1 dimensional quantum field theory},
Commun.~Math.~Phys 175 (1996) 319--336.
\bibitem{EVerlinde} E. P. Verlinde, {\em A note on braid statistics and
the nonabelian Aharonov-Bohm effect}, Proceedings of  the International
Colloquium on  Modern Quantum Field Theory, Bombay, World Scientific
1990.
\bibitem{IZ} C.~Itzykson and J.-B.~Zuber, {\em Quantum Field Theory},
McGraw Hill, Singapore, 1985.
\bibitem{AT} A.~Achucarro and P.~Townsend, {\em  A Chern--Simons
 action for three-dimensional anti-de Sitter supergravity
 theories,}  Phys.~Lett. {\bf B180} (1986)  85--100.
\bibitem{Witten1} E.~Witten, {\em  2+1 dimensional gravity as an exactly
 soluble system,} Nucl.~Phys.  {\bf B311} (1988) 46--78.
\bibitem{Sharpe} R.~W.~Sharpe, {\em Differential Geometry},
Springer Verlag, New York, { 1996}.
\bibitem{AB} M.~Atiyah and R.~Bott, {\em The Yang-Mills equations over
Riemann surfaces,} Phil.~Trans.~Roy.~Soc. London Ser. A {\bf 308}
(1983) 523--615.
\bibitem{Atiyah} M.~Atiyah, {\em The geometry and physics of knots,}
Cambridge University Press, Cambridge, {1990}.
\bibitem{KM}
T.~H.~Koornwinder and N.~M.~Muller.
{\em The quantum double of a (locally) compact group,}
J.~Lie Theory {\bf  7} (1997) 33--52; {\bf 8} (1998) 187 (erratum).
\bibitem{KBM} T.~H.~Koornwinder, F.~A.~Bais and N.~M.~Muller,
{\em Tensor Product Representations of the Quantum Double of a Compact
Group,} Commun.~Math.~Phys. {\bf 198} (1998) 157--186.
\bibitem{SGJ} Ph. de Sousa Gerbert and  R. Jackiw, {\em Classical and quantum
scattering on a spinning cone}, Commun. Math. Phys.{\bf 124} (1988) 229--260.
\bibitem{Gott} J.~R.~Gott, {\em Closed timelike curves produced by
pairs of moving cosmic strings: exact solutions}, Phys.~Rev.~Lett.
{\bf 66} (1991) 1126--1129.
\bibitem{WC} F.~Wilczek and L.~Cornalba, {\em Mass splittings from
symmetry observations}, Phys.~Lett. {\bf  B411} (1997) 112--116.
\bibitem{SG} Ph. de Sousa Gerbert, {\em On spin and (quantum)
gravity in 2+1 dimensions}, Nucl.~Phys. {\bf B346} (1990) 440--472.
\bibitem{CCV1} A.~Cappelli, M.~Ciafaloni, P.~Valtancoli, {\em
Classical scattering in 2+1 gravity with N spinning sources}, 
Phys.~Lett. {\bf B273} (1991) 431--437.
\bibitem{CCV2} A.~Cappelli, M.~Ciafaloni, P.~Valtancoli, {\em
Classical scattering in 2+1 gravity with N point sources}, 
Nucl.~Phys. {\bf 369} (1992) 669--706.
\bibitem{KMVW} K.~Koehler, F.~Mansouri, C.~Vaz and L.~Witten,
{\em Two-particle scattering in the Chern Simons Witten theory
of gravity in 2+1 dimensions}, Nucl.~Phys. {\bf B348} (1991) 373-389.
\bibitem{VW} C.~Vaz and L.~Witten, {\em On the multiparticle phase
space in Chern-Simons gravity}, Nucl.~Phys. {\bf B368} (1992) 509--526.
\bibitem{KO} D.~Kabat and M.~E.~Ortiz, {\em Canonical quantization and 
braid invariance of (2+1)-dimensional gravity coupled to point particles},
Phys.Rev. {\bf D 49} (1994) 1648--1688.
\bibitem{BuRo} 
E.~Buffenoir and Ph.~Roche, {\em Harmonic analysis on the quantum
Lorentz group,} Commun.~Math.~Phys. {\bf 207} (1999) 499--555.
\bibitem{BNR} E~Buffenoir, K~Noui and  P~Roche, {\em 
Hamiltonian Quantization of Chern-Simons theory with SL(2,C) Group}, 
 {\tt hep-th/0202121}.
\bibitem{Bargman}
V.~Bargman, {\em Irreducible unitary representations of the Lorentz group},
Ann.~Math. {\bf 48} (1947) 568--640.
\bibitem{Varadarajan} V.~S.~Varadarajan, {\em Lie groups, Lie algebras
and their representations}, Springer Verlag, New York, 1984.
\bibitem{Muller}
N.~Muller, {\em Topological interactions and quantum double symmetries,}
Ph.D. dissertation, University of Amsterdam, { 1998}.



\end{thebibliography}
\end{document}